\documentclass[3p,review]{elsarticle} %for arXiv, submit
\usepackage{lineno,hyperref}

\usepackage{amsmath,amssymb}
\usepackage{subfig}

\newcommand{\Fig}[1]{#1}
\newcommand{\English}[1]{#1}
\newcommand{\newclearpage}{\clearpage}

\usepackage{xspace}
\newcommand{\Xe}{Xe+CH$_4$\xspace}
\newcommand{\Nuc}[4]{$^{#1}_{\;\;#2}\rm{#3}^{#4}$\xspace}
\newcommand{\Ncc}{$\overline{N}_{cc}$\xspace}
\newcommand{\Up}{$^{238}\rm{U}$\xspace} %primary uranium beam 238U
\newcommand{\Uppp}{$^{238}\rm{U}^{90+,91+}$\xspace}
\newcommand{\Upp}{$^{238}\rm{U}^{90+}$\xspace}
\newcommand{\Brho}{$B\rho$\xspace}
\newcommand{\diameter}{$\varnothing$}
\newcommand{\dray}{$\delta$-ray\xspace}
\newcommand{\drays}{$\delta$-rays\xspace}
\newcommand{\MeVu}{MeV/u\xspace}

\newcommand{\dEdx}{$d E\!/\!d x$\xspace}

\newcommand{\dE}{$\Delta E$\xspace}
\newcommand{\dEfull}{$\Delta E_{full}$\xspace}
\newcommand{\dEBB}{$\Delta E_{colT}$\xspace}
\newcommand{\dEchex}{$\Delta E_{cc}$\xspace}
\newcommand{\dEchexi}{$\Delta E_{cc}(i)$\xspace}

\newcommand{\ERtot}{$R_{total}$\xspace}
\newcommand{\ERcol}{$R_{col}$\xspace}
\newcommand{\ERcc}{$R_{cc}$\xspace}

\newcommand{\Fq}{$F_q$\xspace}

\newcommand{\Rsim}{$\Delta Z_{sim}$\xspace}
\newcommand{\Rexp}{$\Delta Z_{data}$\xspace}
\newcommand{\Rcol}{$\Delta Z_{sim}^{colT}$\xspace}
\newcommand{\Rchex}{$\Delta Z_{sim}^{total}$\xspace}

\newcommand{\Ocol}{$\Omega_{col}$\xspace}
\newcommand{\OBB}{$\Omega_{colT}$\xspace}
\newcommand{\Occ}{$\Omega_{cc}$\xspace}

\newcommand{\Ed}{$E_d$\xspace}
\newcommand{\Em}{$E_m$\xspace}

\newcommand{\Z}{$Z$\xspace}
\newcommand{\Zeq}[1]{$Z=#1$}
\newcommand{\Zg}[1]{$Z>#1$}
\newcommand{\Zl}[1]{$Z<#1$}

\newcommand{\Q}{$Q$\xspace}
\newcommand{\AoQ}{$A/Q$\xspace}

\modulolinenumbers[1]
\journal{Journal of \LaTeX\ Templates}

\begin{document}

\begin{frontmatter}

\title {Xenon-gas ionization chamber to improve particle identification of heavy ion beams with \Zg{70}}

\author[ist1] {Masahiro~Yoshimoto\corref{cor}}
\ead{masahiro.yoshimoto@riken.jp}
\author[ist1]{Naoki~Fukuda}
\author[ist2]{Riku~Matsumura}
\author[ist3]{Daiki~Nishimura}
\author[ist1]{Hideaki~Otsu}
\author[ist1]{Yohei~Shimizu}
\author[ist1]{Toshiyuki~Sumikama}
\author[ist1]{Hiroshi~Suzuki}
\author[ist3]{Hiroyuki~Takahashi}
\author[ist1]{Hiroyuki~Takeda}
\author[ist1]{Junki~Tanaka}
\author[ist1]{Koichi~Yoshida}
\cortext[cor]{Corresponding author}

\address[ist1]{RIKEN Nishina Center, 2-1 Hirosawa, Wako, Saitama 351-0198, Japan}
\address[ist2]{Saitama University, 255 Shimo-Okubo, Sakura-ku, Saitama-shi, Saitama 338-8570 Japan}
\address[ist3]{Tokyo City University, 1-28-1 Tamazutsumi, Setagaya-ku, Tokyo 158-8557, Japan}

\begin{abstract}
\English{
In  conventional ionization chambers (ICs) using P-10 (Ar+CH$_4$) gas, as the atomic number (\Z) of the ion beams increases in the energy region of 200--300~\MeVu, the \Z resolution deteriorates rapidly when \Zg{70}. 
This degradation is attributed to substantial energy loss straggling caused by charge-state fluctuation when the beams traverse a gas medium. 
The energy loss straggling increases when the beams cannot attain charge-state equilibrium in the IC gas. 
In this study, a xenon-based gas (\Xe), exhibiting a sufficiently large charge-state changing cross section, was used in the IC to reach charge-state equilibrium. 
The responses of ICs with P-10 and the xenon-based gases were examined using \Nuc{238}{92}{U}{} beams and cocktail radioactive isotope (RI) beams with \Zeq{40}--90 at the RI Beam Factory (RIBF). 
For \Up beams at 165--344 \MeVu, the P-10 gas IC yielded an energy resolution of 1.9--3.0\% in full width at half maximum (FWHM), which proved inadequate for \Z identification in the uranium region. 
In contrast, the xenon-based gas IC demonstrated a satisfactory energy resolution of 1.4--1.6\%. 
When using cocktail RI beams, a \Z resolution of 1.28 and 0.74 was achieved by the P-10 and the xenon-based gas ICs, respectively, for beams with \Zeq{84}--88 at 200~\MeVu. 
The contrast in \Z resolutions between the P-10 and the xenon-based gas ICs was effectively elucidated by the energy loss straggling model, incorporating collisional straggling and straggling due to charge-state changes in the IC gases. 
The xenon-based gas IC, with more than 3$\sigma$ \Z separation across a broad \Z range (\Zeq{40}--90), emerged as a practical solution for \Z identification of heavy ion beams.
}
\end{abstract}


\end{frontmatter}

% \linenumbers
\section{Introduction}

\English{
Radioactive isotope (RI) beams are appropriate for studying short-lived unstable nuclei. 
These beams are produced through in-flight fission or projectile fragmentation of energetic primary beams comprising stable heavy ions. 
Objective RIs are subsequently separated and identified using in-flight RI-beam separators; these RIs are then used for experimental purposes.
Since 2007, the RI Beam Factory (RIBF) has been a world-leading facility producing various RI beams at 200--300~\MeVu~\cite{Yano2007NIMB,Kubo2012PTEP}.
Primary beams of \Up, \Nuc{124}{}{Xe}{}, \Nuc{78}{}{Kr}{}, \Nuc{70}{}{Zn}{}, \Nuc{48}{}{Ca}{}, etc. are accelerated to 345~\MeVu by a Superconducting Ring Cyclotron (SRC)~\cite{Yano2007NIMB}; subsequently, RI beams are produced, separated, and identified in the BigRIPS separator~\cite{Kubo2012PTEP}.
A total of 171 new isotopes with a wide range of atomic numbers (\Z) from 11 to 68~\cite{NewRIBigRIPS0}--\cite{NewRIBigRIPSd} have been discovered at RIBF; moreover, the neutron dripline for F and Ne has been determined~\cite{DSAhn2019PRL}.
Additionally, many other experiments based on inverse kinematics, such as Refs.~\cite{Crawford2019PRL,Taniuchi2019Nature,Bagchi2020PRL} have been conducted.
}

\English{
The identification of RI beams involves deducing their atomic numbers (\Z) and mass-to-charge ratios (\AoQ) on an event-by-event basis. 
This is achieved through a TOF--\Brho--\dE (TOF; time-of-flight, \Brho; magnetic rigidity, and \dE; energy loss) method utilizing beamline detectors in BigRIPS~\cite{Fukuda2013NIMB}. 
TOF is measured using plastic scintillators; \Brho values are determined through trajectory reconstruction from tracks measured using position-sensitive parallel plate avalanche counters (PPACs)~\cite{Kumagai2013NIMB}; and \dE is measured using a multi-sampling ionization chamber (MUSIC, hereafter IC)~\cite{Christie1987NIMA,Kimura2005NIMA}.
This particle-identification (PID) method has been successfully utilized to produce RI beams in the \Zl{70} region at RIBF. 
Recently, heavier RI beams with \Zg{70} have been produced using the projectile fragmentation of the \Up primary beam, giving rise to the following issues ~\cite{Sumikama2020NIMB,Inabe2015APR,Fukuda2021APR}:
(1) Neutron-rich RI beams face challenges in achieving sufficient separation from lower-\Z contaminants produced through secondary reactions in energy degraders.
(2) Yields of RI beams experience a drastic reduction during the separation and particle identification (PID) processes, which involve alterations in charge states at the beamline detectors and the energy degraders.
(3) A higher \AoQ  resolution is essential due to the denser \AoQ spectrum resulting from larger \Q and the presence of multiple charge states.
(4) \Z resolution deteriorates in the IC with the widely-used P-10 (90\% Ar + 10\% CH$_4$) gas.
At RIBF, Sumikama {\it et al.} have partly overcome the PID issues by transporting only He-like ions in \Nuc{208}{86}{Rn}{84+} setting at 185~\MeVu~\cite{Sumikama2020NIMB}.
Single charge-state beams near $A/Q=2.5$ require a separation of only $\delta Z=2$ difference in the two-dimensional PID spectrum of \Z vs. \AoQ; hence, even a P-10 gas IC with a low \Z resolution of approximately one in FWHM can be used for \Z identification.
However, other RI beams with \AoQ deviating from 2.5 and/or those with multiple charge states require a separation of $\delta Z = 1$ difference. 
Thus, our focus is on enhancing the \Z resolution itself to effectively handle common heavy RI beams.
}

\English{
The \Z resolution is affected by the variation of the energy loss, called \dE straggling, in the IC gas.
For heavy ion beams, a drastic increase in the \dE straggling has been observed, which is caused by changes in the charge state, namely the electron-capture and electron-loss processes~\cite{Weick2000PRL}.
To address this concern, a thick P-10 gas IC with 800 mg/cm$^2$ was utilized at GSI \cite{Audouin2005NIMA} for \Z identification of heavy RI beams around \Zeq{80} with 300~\MeVu. 
This method augments the number of charge-state changes, approaching charge-state equilibrium, which is the asymptotic average charge state attained after many charge-state changes. 
Although this approach yields a satisfactory \Z resolution, there is an energy loss of approximately 80~\MeVu in the thick IC, rendering it impractical for supplying RI beams for direct reaction experiments at RIBF. 
To increase the number of charge-state changes under the same energy loss, a xenon-based gas IC was proposed, offering a charge-state changing cross section approximately one order of magnitude larger than that of ICs with P-10 gas~\cite{Sumikama2023APR}.
}

\English{
This study  compares the \Z resolution of the newly-proposed xenon-based gas IC for \Zeq{40}--92 beams to that of the conventional P-10 gas IC.
Section~\ref{sec2} explains the effect of charge-state changes on energy resolution and \Z resolution.
Section~\ref{sec3} describes the structure of the proposed IC, electronics, and IC gases.
Section~\ref{sec4} summarizes the \Up beams and cocktail RI beams with \Zeq{40}--90 used to evaluate the energy resolution and \Z resolution, respectively.
The results for the energy resolution and \Z resolution are presented in Section~\ref{sec5} and a discussion on the \Z resolution using simulations of charge-state fluctuations is presented in Section~\ref{sec6}.
Each resolution is given in full width at half maximum (FWHM) throughout this paper unless specified otherwise.
}

\section{Effect of charge-state changes on energy resolution and \Z resolution\label{sec2}}

\English{
The degraded \Z resolution for heavy ion beams at 200--300~\MeVu is due to enhanced \dE straggling caused by charge-state changes~\cite{Weick2000PRL}.
The mean energy-loss rate (\dEdx) is a function of \Q not \Z; thus, the \dE value depends on the charge state of each event in the medium.
}

\English{
For fully-stripped ions, \dEdx follows the Bethe equation using \Z and velocity ($\beta$) of the ions as
\begin{linenomath*}\begin{equation}\label{eq:bethe}
	\frac{d E}{d x} = Z^2 \frac{4 \pi e^4}{m \beta^2c^2}n_e\left(\frac{1}{2} \ln \frac{E_m^2}{I^2} - \beta^2 +\Delta_{cor}\right),
\end{equation}\end{linenomath*}
where $e$ is the elementary charge, $m$ is the electron mass, $c$ is the speed of light, $n_e$ is the electron density, $I$ is the mean excitation energy, \Em is the maximum energy transfer to a target electron in a single collision of the ion and a material atom, and $\Delta_{cor}$ is the various corrections term~\cite{Lindhard1996PRA}.
The collisional straggling (\Ocol) in energy loss is expressed as follows:
\begin{linenomath*}\begin{equation}\label{eq:omega}
	\frac{d\Omega_{col} ^2}{d x} =Z^2 \frac{2 \pi e^4}{m \beta^2c^2} n_e E_m X_{LS},
\end{equation}\end{linenomath*}
where $X_{LS}$ is the correction term by Lindhard and S\o rensen~\cite{Lindhard1996PRA}.
The collisional straggling is mainly caused by stochastic fluctuations in the number and energy of released high-energy electrons (\drays), which are emitted when the energy transfer is large.
Because the IC gas at BigRIPS is sufficiently thin to use a constant $\beta$ assumption, the mean energy loss of the fully-stripped ions (\dEfull) is expressed as
\begin{linenomath*}\begin{equation}\label{eq:efull}
  \Delta E_{full} = \xi(\beta) L Z^2,
\end{equation}\end{linenomath*}
where $\xi (\beta)$ is obtained from the right-hand side of Eq.~\ref{eq:bethe} excluding $Z^2$, and $L$ is the gas thickness.
The \Z value is obtained by solving Eq.~\ref{eq:efull} and the relative uncertainty of \Z is half of the fluctuation of \dEfull.
}

\English{
In the case of partially-stripped ions, \dEdx is derived from Eq.~\ref{eq:bethe}, in which \Z is replaced by \Q ($<$\Z).
The energy loss of the $i$-th ion (event no.~$i$) with charge-state changes (\dEchexi) is expressed as follows:
\begin{linenomath*}\begin{equation}\label{eq:chex1}
  \Delta E_{cc}(i) = \xi (\beta) \sum_{j=0}^{N_{cc}(i)} L_j(i) q_j^2,
\end{equation}\end{linenomath*}
where $N_{cc}(i)$ is the number of charge-state changes of the $i$-th ion, and $L_j(i)$ is the thickness that the ion with charge $q_j$ after $j$-times changes passes through in the IC gas.
Moreover, \dEchexi is given with the ratio $F_q(i)$ of the combined thickness that the ion with charge $q$ passes to the gas thickness $L$ via the following expression:
\begin{linenomath*}\begin{equation}\label{eq:chex2}
  \Delta E_{cc}(i) = \xi (\beta) L \sum_{q=1}^Z F_q (i) q^2,
\end{equation}\end{linenomath*}
where $\sum F_q$ is equal to one.
Here, the mean energy loss of all ions (\dEchex) is less than \dEfull.
The fluctuation of \dEchex with each event, called charge-exchange straggling (\Occ), occurs when the charge-state distribution broadens and \Fq varies with each event.
The fluctuation of \Fq depends on $N_{cc}$.
The average $N_{cc}$ (\Ncc) is expressed as follows:
\begin{linenomath*}\begin{equation}\label{eq:ncc}
  \overline{N}_{cc} = \sum_{q=1}^{Z} \overline{F}_q\frac{L}{\lambda_q},
\end{equation}\end{linenomath*}
where $\lambda_q$ is the mean free path of an ion with charge $q$ and $\overline{F}_q$ is the mean existence probability of ions with charge $q$. 
In general, the charge-exchange straggling becomes large when ions prefer being partially stripped in the gas and \Ncc is inadequate to attain charge-state equilibrium.
}

\begin{figure}[h]
\centering
\Fig{\includegraphics[width=0.7\textwidth]{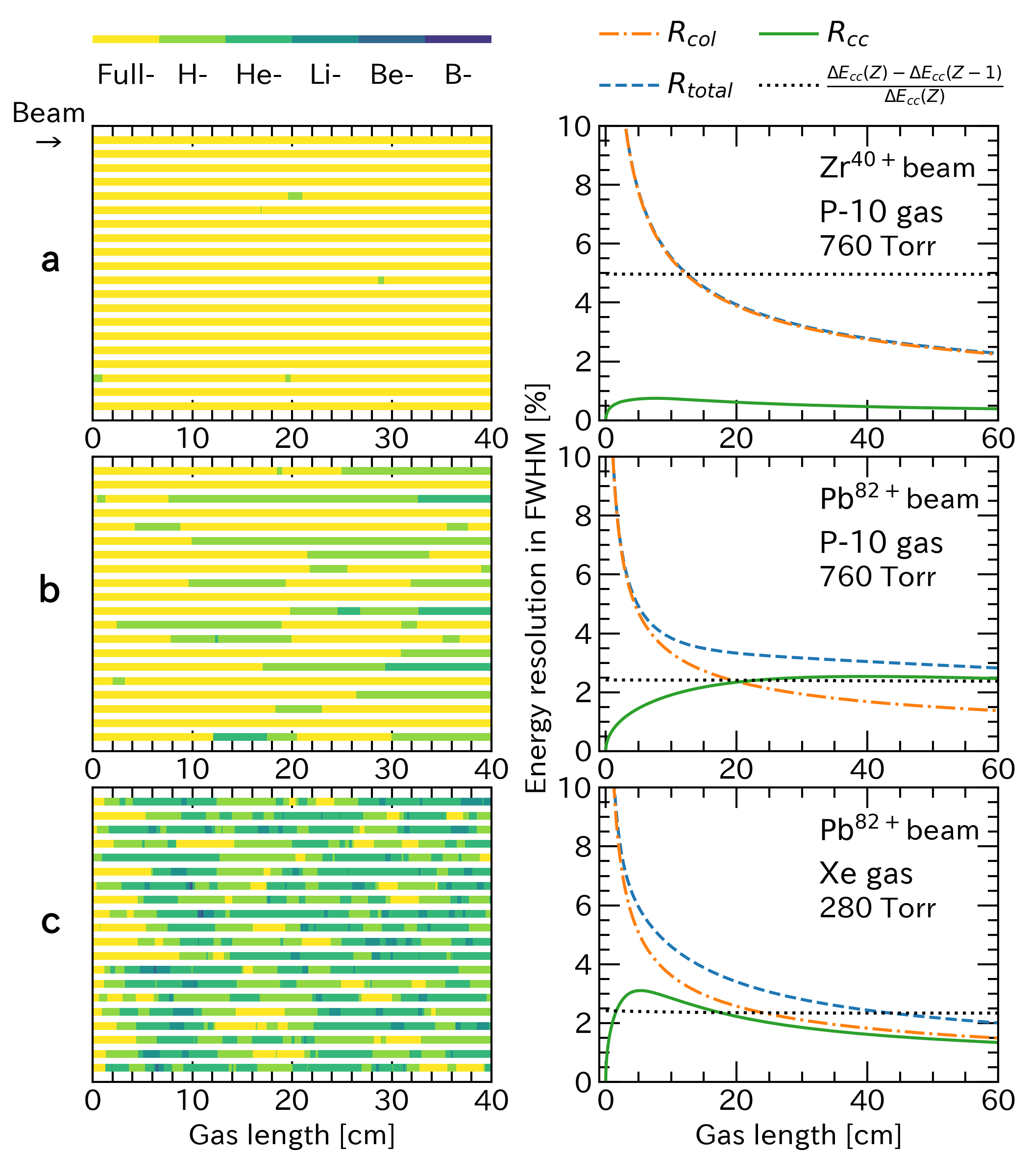}}
\caption{
		Simulations for charge-state fluctuation and energy resolution of (a)\Nuc{100}{}{Zr}{40+} into P-10 gas, (b)\Nuc{205}{}{Pb}{82+} into P-10 gas, and (c)\Nuc{205}{}{Pb}{82+} into Xe gas.
		The beam energy is 250~\MeVu.
		The pressure of the P-10 gas is 760~Torr, while that of the Xe gas is set at 280~Torr to achieve an equivalent \dE.
		The horizontal bars in the left figures display the generated 20 events using the Monte Carlo method, and the color changes indicate the charge-state changes in the gas.
		The energy resolutions in the right figures were defined as $R=2.355\Omega/\Delta E_{cc}$.
		The energy resolutions, \ERcol and \ERcc were deduced with simulated collisional straggling using Eq.~\ref{eq:omega} and charge-exchange straggling using Eq.~\ref{eq:chex2}, respectively.
		The energy resolution with total \dE straggling, \ERtot was estimated by adding the collisional and charge-exchange straggling in quadrature. 
		The mean energy loss, $\Delta E_{cc}$ was estimated using the CATIMA energy loss calculator~\cite{CATIMA} and Eq.~\ref{eq:chex2}.
		}
\label{fig-MC_P10Xe}
\end{figure}

\English{
To elucidate the collisional and charge-exchange straggling, Fig.~\ref{fig-MC_P10Xe} illustrates examples of simulated charge-state fluctuations and energy resolution ($R=2.355\Omega/\Delta E_{cc}$) at 250~\MeVu.
The charge-state fluctuations were simulated using a dedicated Monte Carlo code, with charge-state changing cross sections taken from the GLOBAL code~\cite{GLOBAL}.
The collisional straggling (\Ocol) is calculated using Eq.~\ref{eq:omega}, whereas \dEchex and \Occ are calculated using Eq.~\ref{eq:chex2}.
Figure~\ref{fig-MC_P10Xe} illustrates results for \Nuc{100}{40}{Zr}{40+} beam in P-10 gas (a), \Nuc{205}{82}{Pb}{82+} beam in P-10 gas (b), and \Nuc{205}{82}{Pb}{82+} beam in Xe gas (c). 
The horizontal bars in the left figures display the charge-state changes in the gas by color.
The right figures show the energy resolution as a function of gas length.
The orange dashed-dotted lines indicate the energy resolution of the collisional straggling (\ERcol), and the green solid lines indicate the energy resolution of the charge-exchange straggling (\ERcc).
The blue dashed lines indicate the energy resolution of the total \dE straggling (\ERtot), which was calculated by adding \Ocol and \Occ in quadrature.
The black dotted lines indicate the relative \dEchex difference between \Z and $Z-1$ as follows:
\begin{linenomath*}\begin{equation}\label{eq:dediff}
	\frac{\Delta E_{cc}(Z)-\Delta E_{cc}(Z-1)}{\Delta E_{cc}(Z)},
\end{equation}\end{linenomath*}
which represents the required energy resolution to separate \Z at 2.355$\sigma$.
}

\English{
When the \Nuc{}{}{Zr}{40+} beam is injected into the 760-Torr P-10 gas as shown in Fig.~\ref{fig-MC_P10Xe}(a), \ERcc is sufficiently small. Thus, \ERtot is dominated by the collisional straggling. 
Here, \ERcol decreases as a function of $1/\sqrt{L}$ from the relationship between Eqs.~\ref{eq:chex2} and \ref{eq:omega}.
The charge-exchange straggling is small because the Zr ions prefer being fully stripped to being partially stripped.
Most ions remain fully stripped (yellow) throughout the gas.
Some ions become H-like ions (light green); however, these H-like ions immediately revert to fully-stripped ions within short flight distances, thereby rendering the charge-state distribution narrow.
As presented in a previous publication~\cite{Sato2013JApplPhys}, the energy loss resolution obtained by the P-10 gas ICs translates to a sufficient \Z resolution for ions with \Zl{50} at 200--300\MeVu.
}

\English{
When the \Nuc{}{}{Pb}{82+} beam is injected into the 760-Torr P-10 gas as shown in Fig.~\ref{fig-MC_P10Xe}(b), \ERcc affects \ERtot.
The \Nuc{}{}{Pb}{82+} ions tend to capture electrons in the P-10 gas. Thus, many ions change to H-like or He-like (green) in the gas.
Since the $\lambda$ of the partially-stripped Pb ions is longer than that of the Zr ions, the H-like and He-like Pb ions pass through the gas as partially-stripped ions with longer flight distances. 
Only a few ions remain fully stripped until the end of their trajectory through the gas.
The charge-state distribution widens, and \Fq varies with each event, thereby resulting in large \ERcc.
The inadequate \Z resolution for ions with \Zeq{80}--90 using the P-10 gas has been observed in earlier studies~\cite{Sumikama2020NIMB,Inabe2015APR,Fukuda2021APR} at RIBF. 
}

\English{
Figure~\ref{fig-MC_P10Xe}(c) shows a case in which the \Nuc{}{}{Pb}{82+} beam is injected into the Xe gas.
The Xe gas pressure is adjusted to be 280~Torr to induce an energy loss equivalent to that of the P-10 gas. Thus, \ERcol is the same as the result obtained using the P-10 gas.
The charge-state distribution in the Xe gas is broader than that in the P-10 gas.
The charge-state changes occur many times because the $\lambda$ of partially-stripped Pb ions in the Xe gas is approximately an order of magnitude shorter than that in the P-10 gas.
The simulated \Ncc is 23 in the 40-cm Xe gas, which is sufficient to reach charge-state equilibrium, while \Ncc of 2.2 in the 40-cm P-10 gas is insufficient.
Although the charge-state distribution in the Xe gas becomes broader than that in the P-10 gas, the fluctuation of \Fq is smaller. Hence, \ERcc is reduced as $L$ increases.
The \Z separation is expected to be possible with the Xe gas longer than 44~cm.
}

\English{
The \ERcc peak positions of the \Nuc{}{}{Pb}{82+} beam are $L$=40~cm for the P-10 gas and $L$=5~cm for the Xe gas. Both have an \Ncc value of approximately 2.
When \Ncc is small, \ERcc increases because the effect of the broadening charge-state distribution as \Ncc increases is dominant.
When \Ncc exceeds 2, \ERcc decreases because the effect of averaging the charge-state fraction \Fq becomes dominant.
The \ERcc value is proportional to $\sqrt{L}$ at \Ncc~$<2$ and $1/\sqrt{L}$ at \Ncc~$>2$, similar to \ERcol.
A choice to use short ICs with the P-10 gas to reduce \Ncc and \ERcc with the \Nuc{}{}{Pb}{82+} beam does not work because \ERcol becomes large.
Another choice to use long ICs is also difficult at RIBF.
To achieve the required energy resolution for the \Nuc{}{}{Pb}{82+} beam, 120-cm gas is required; however, this is not feasible with the BigRIPS layout.
}

\newclearpage
\section{Development of the ionization chamber\label{sec3}}

\subsection{Structure}

\begin{figure*}[h]
\centering
\Fig{\includegraphics[width=0.7\textwidth]{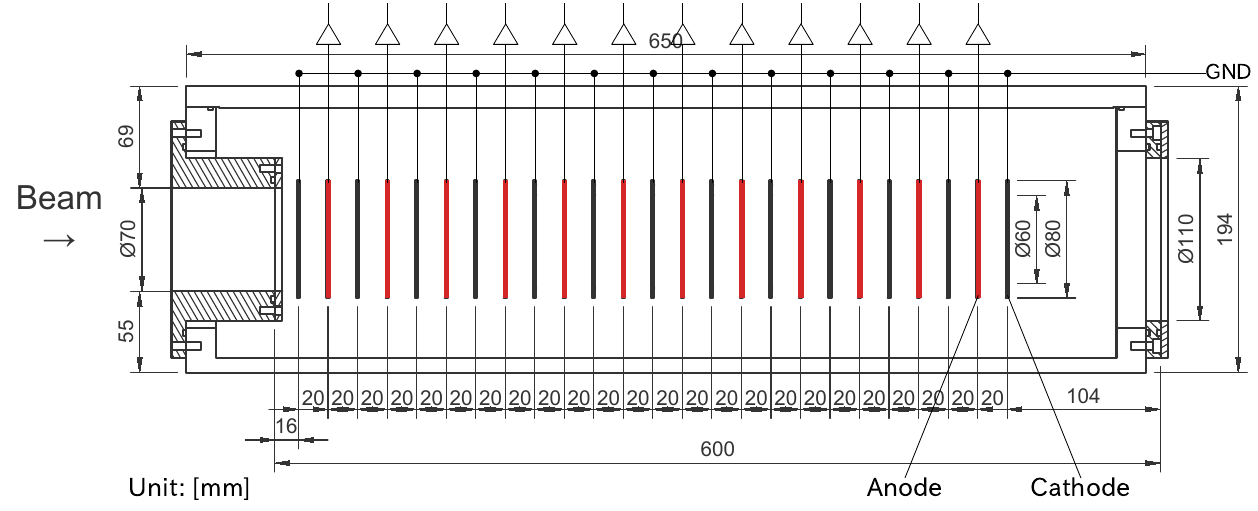}}
	\caption{
	Cross-sectional view of the newly-developed multi-sampling ionization chamber. 
	The red and black bold lines show 12 anodes and 13 cathodes, respectively, whose inner diameters are 6.0~cm.
	The active gas length is 48~cm.
	}
\label{fig-IC_cs}
\end{figure*}

\English{
A cross-sectional view of the newly-developed IC is shown in Fig.~\ref{fig-IC_cs}.
This IC comprises 25 electrode planes aligned perpendicular to the beam axis.
A total of 12 anodes and 13 cathodes are arranged alternately with 2.0-cm intervals. 
The active gas length is 48~cm.
The electrodes are made of 2.5-$\mu$m thick aluminized Mylar films, supported by pairs of 1-mm thick aluminum rings with inner and outer diameters of 6.0 and 8.0~cm, respectively.
The effective area is a circle with \diameter 6~cm as uniform parallel electric fields are formed in the volume inside the inner diameters.
This area is reduced from that of previous ICs~\cite{Sato2013JApplPhys,Sato2012APR} comprising a circle of \diameter 23.2~cm or a rectangle of 26$\times$17~cm$^2$ because smaller electrodes may reduce the collisional straggling by exporting energy deposition of the high-energy \drays outside the active detection volume~\cite{Pfutzner1994NIMB}.
In addition, the smaller electrodes reduce capacitive noises.
These electrodes are enclosed in an airtight aluminum chamber filled with IC gases.
Upstream and downstream vacuum separation windows are 77-$\mu$m and 125-$\mu$m thick Kapton sheets, respectively, and their sizes are \diameter 7~cm and \diameter 11~cm, respectively.
The distance between the windows is 60~cm.
The inactive gas length from the upstream window to the first electrode is 1.6~cm, which is reduced from the 9.6~cm or 7.0~cm of previous ICs.
}

\subsection{Electronics}

\English{
Ionization electrons originating from gas atoms are collected by the IC anodes under parallel electric fields that do not cause electron avalanches.
A total of 12 output signals proportional to the energy deposit to each active volume are multi-sampled.
Each signal was amplified by a preamplifier with 7-$\mu$s decay time and a shaping amplifier with 1-$\mu$s shaping time, and then processed by a 14-bit ADC.
All 13 cathodes were grounded to the airtight aluminum chamber with copper wires.
The preamplifiers were operated in vacuum and attached directly to the outside of the airtight chamber to reduce ground noise and stray capacitance.
The 12 ADC values were geometrically averaged in the analysis to obtain \dE.
}

\subsection{Gases}

\English{
The IC is filled with either the commonly-used P-10 gas (90\% Ar + 10\% CH$_4$) or the newly-proposed xenon-based gas (70\% Xe + 30\% CH$_4$). 
The gas pressures are controlled to be 620~Torr for both gases by the constraint of the gas-flow device. 
In the case of the P-10 gas, the active gas thickness  is 61.2 mg/cm$^2$, and a voltage of +400~V is supplied to the anodes to maximize the electron drift velocity, estimated to be 4.9~cm/$\mu$s using Garfield++~\cite{Garfield++}. 
For the \Xe gas, the gas mixture ratio is set at 7:3, and the active gas thickness is 158~mg/cm$^2$. 
Although a higher percentage of xenon gas is desirable to take advantage of xenon properties on charge-state changing, it slows down the electron drift velocity, resulting in a broad signal waveform from the electrodes. 
This broadening causes an increase in pile-up events under high-rate beam conditions. 
Hence, a gas mixture ratio of 7:3 is adapted to maintain the same electron drift velocity as the P-10 case. 
A voltage of +800~V is supplied for the \Xe gas in our experiment. 
Although a voltage of +1200~V was initially required to match the velocity of the P-10 gas, unforeseen discharge problems necessitated the use of the lower voltage, resulting in a calculated electron drift velocity of 3.5~cm/$\mu$s with the +800~V voltage.
}

\English{
The mean energy required to produce an electron-ion pair is 26~eV for the P-10 gas and 23~eV for the \Xe gas.
The number of initial electrons in the \Xe gas increases by approximately 10\% under the same energy loss condition, however, this increase is negligible for the improvement of the energy resolution.
The energy loss exceeds 100~MeV in the IC gases for heavy RI beams with a typical energy of RIBF.
The number of initial electrons is more than $10^6$, and thus its statistical fluctuation is less than 0.1\%, which is sufficiently small to affect the energy resolution.

The energy loss of the \Xe gas IC is approximately two times that of the P-10 gas IC under the same gas pressure used for our measurements.
Because the estimated energy resolution with the collisional straggling (\ERcol) of the \Xe gas IC is 0.7($=1/\sqrt{2}$) times that of the P-10 gas IC, the \Xe gas IC could achieve superior energy resolution and \Z resolution in a low-\Z region where \Ocol is dominant and \Occ is negligible.
However, \Z resolution did not improve in the low-\Z region in our experiment.
The cause of this phenomenon could be high-energy \drays escaping from the active detection volume, which is discussed in Section~\ref{sec6}.
}

\newclearpage
\section{Experiment\label{sec4}}

\English{
The required range of RI beams for ICs in the BigRIPS separator spans up to \Zeq{92}, encompassing energies of 150--300~\MeVu with multiple charge states.
To evaluate the P-10 and \Xe gas ICs, \Nuc{238}{92}{U}{90+, 91+}beams with an energy range of 165--344~\MeVu and cocktail RI beams with a wide \Z range from 40 to 90 at 200--250~\MeVu were injected into the ICs.
}

\subsection{$^{238}\rm{U}$ beams}

\begin{figure*}[htbp]
	\centering
	\Fig{\includegraphics[width=0.9\textwidth]{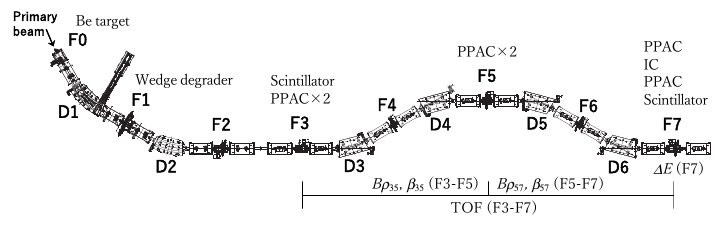}}
	\caption{
		Layout of BigRIPS separator including focal planes, F0 to F7, and dipoles, D1 to D6.
		The combination of materials and detectors installed at each focal plane varies depending on the beam settings.
		The time of flight (TOF) is measured with the F3 and F7 plastic scintillators.
		Magnetic rigidity, $B\rho_{35}$ and $B\rho_{57}$, are determined through trajectory reconstruction with PPACs at F3, F5, and F7.
		Velocity, $\beta_{35}$ and $\beta_{57}$ are deduced by TOF and \Brho values.
		The energy loss, \dE is measured with the IC at F7. 
	}
	\label{fig-layout}
\end{figure*}

\English{
The \Up beams were provided at energies of 344, 252, and 165~\MeVu with $Q\!=$90+ and 91+.
These charge states are widely used for heavy ion beams at a typical energy of 250~\MeVu at RIBF. 
The three energies were achieved by degrading the primary beam of 346.6-\MeVu \Nuc{238}{}{U}{86+} through a 0.2-mm thick plastic scintillator at BigRIPS focal plane F3, a 5-mm thick Be target at F0, and a 9-mm thick Be target at F0, respectively (see Fig.~\ref{fig-layout} for BigRIPS layout). 
These materials also served as electron strippers.
The 90+ or 91+ ions were selectively filtered by slits at the momentum dispersive focal plane, F5. 
The ICs were positioned in the F7 chamber. 
For the determination of beam timings and trajectories into the ICs, a 0.1-mm thick plastic scintillator and two PPACs were installed in the F7 chamber. 
The sequence of detectors from upstream to downstream comprised the first PPAC, IC, second PPAC, and scintillator. 
To mitigate pile-up events, the beam intensity at F7 was regulated to approximately 1~kHz.
}

\subsection{Cocktail RI beams of $Z=$40--90}

\English{
The cocktail RI beams spanning \Zeq{40}--90 were produced by the primary beam, \Nuc{238}{}{U}{86+}, colliding with a 4-mm thick Be target at F0. 
The beams were then separated using the first dipole (D1) with a \Brho of 6.1712~Tm and a 2-mm thick Al wedge energy degrader at F1. 
The \Brho settings after F1 were adjusted for the \Nuc{210}{}{Rn}{84+} beam, limiting the momentum width to $\pm$0.05\% by narrowing the F1 slit to $\pm$1~mm. 
For PID, the TOF between F3 and F7 was measured using 0.2-mm and 0.1-mm thick plastic scintillators, respectively. 
The \Brho values were determined through trajectory reconstruction with PPACs installed at F3, F5, and F7. 
The \AoQ and $\beta$ values of each particle were deduced from the TOF and \Brho values~\cite{Fukuda2013NIMB}.
Considering partially-stripped ions, effective \Q ($Q_{eff}=\sqrt{\sum F_q q^2}\le Z$) instead of \Z is derived from \dE and $\beta$ by using Eq.~\ref{eq:chex2}.
In practice, the measured $Q_{eff}$ was adjusted to \Z using third-order polynomials.
PID was further confirmed by detecting delayed $\gamma$ rays emitted from the short-lived isomeric states of \Nuc{138}{}{Ba}{}, \Nuc{210}{}{Rn}{}, and \Nuc{215}{}{Ra}{}~\cite{NuDat}. 
The beam intensity was regulated to approximately 1~kHz at F7, consistent with one of the \Up beams.
}

\section{Results\label{sec5}}

\subsection{$^{238}\rm{U}$ beams\label{sec-ubeam}}

\begin{figure*}[htbp]
	\centering
	\Fig{\includegraphics[width=0.6\textwidth]{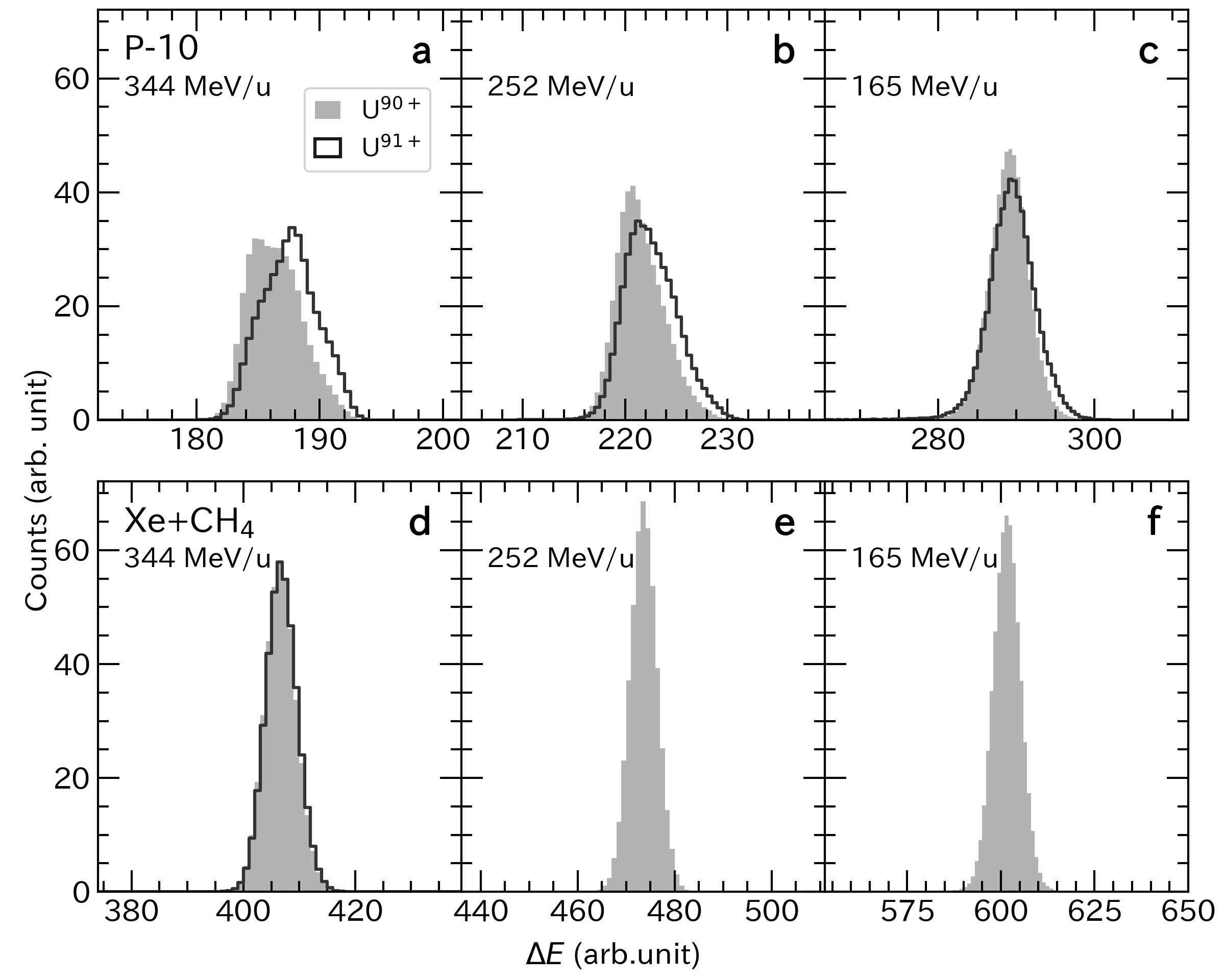}}
	\caption{
		Energy  loss (\dE) distributions of \Up beams at 344, 252, and 165~\MeVu measured by the P-10 and \Xe gas ICs.
		The filled and open histograms represent the \dE distributions of incident 90+ and 91+ ions, respectively.
		The horizontal axes of all spectra are scaled from 92\% to 108\% of the mean \dE for comparison.
		The areas of all \dE distributions are normalized to be equal.
	}
	\label{fig-U_dE}
\end{figure*}

\English{
The energy-loss (\dE) distributions of the \Uppp beams at 344, 252, and 165~\MeVu measured by the P-10 and \Xe gas ICs are illustrated in Fig.~\ref{fig-U_dE}. 
The filled and open histograms depict the \dE distributions of 90+ and 91+ ions, respectively. 
Table~\ref{tab-U_dE} provides a summary of the mean \dE values, mean \dE differences between 90+ and 91+, energy resolutions, and the average number of charge-state changes in the gas (\Ncc) at the equilibrium charge state calculated using Eq.~\ref{eq:ncc}.
}

\English{
The \dE distributions of the \Uppp beams at 344~\MeVu measured by the P-10 gas IC are presented in Fig.~\ref{fig-U_dE}(a). 
A noticeable difference in the distribution between 90+ and 91+ is observed, where the mean \dE of 91+ is 0.62\% larger than that of 90+. 
The total thickness of the active gas and upstream substances is considerably thinner than the equilibrium thickness for 344-\MeVu \Up beams, resulting in a mean \dE dependence on the incident charge states. 
An asymmetric \dE distribution for 90+ and a more symmetric distribution  for 91+ were observed. 
This difference is caused by the gap between the incident charge states and the mean charge of 91.0+ at equilibrium in the gas.
The high energy shoulder of the 90+ distribution is formed by the incident 90+ ions that were changed to 91+ in the gas.
The 91+ distribution exhibits tails on both sides due to a portion of the incident 91+ ions converting to 90+ or 92+ ions. 
The energy resolutions are 3.0\% for the 90+ and 2.9\% for the 91+ ions. 
The required energy resolution for separating uranium (\Zeq{92}) from protactinium (\Zeq{91}) at 165--344~\MeVu is 2.1\% calculated using Eq.~\ref{eq:dediff}. 
Considering the substantial mean \dE difference of 0.62\% and the inadequate energy resolutions of 3\%, \Z separation with the P-10 gas IC is unattainable in the uranium region at 344~\MeVu.
}

\English{
The \dE distributions of the \Uppp beams at 252 and 165~\MeVu, respectively, measured by the P-10 gas IC are presented in Figs.~\ref{fig-U_dE}(b) and (c). 
As the energy decreases, the mean \dE difference between 90+ and 91+ decreases, the energy resolutions improve, and the distributions of the two charge states become more similar and less asymmetric. 
The improvement at lower energies is attributed to an increase in mean \dE and a decrease in charge-exchange straggling due to the increase in \Ncc. 
Even with the energy reduced to 165~\MeVu, \Z separation remains unattainable due to insufficient energy resolution and residual mean \dE difference.
}

\English{
Figure~\ref{fig-U_dE}(d) displays the \dE distributions of the \Uppp beams at 344~\MeVu measured by the \Xe gas IC. 
The mean \dE difference is 0.03\%, which is negligible for the required energy resolution in the uranium region. 
The energy resolutions of 90+ and 91+ are both 1.6\%, markedly better than those obtained by the P-10 gas IC. 
The distributions are nearly identical and symmetrical. 
Although the charge states near the upstream of the active gas  depend on the incident charge states, the charge distribution quickly reaches equilibrium in the \Xe gas, eliminating the incident charge-state dependence in the \dE distribution. 
The \Xe gas IC exhibits a negligible mean \dE difference and superior energy resolutions compared to the 2.1\% requirement, enabling \Z separation in the uranium region even at 344~\MeVu.
}

\English{
The \dE distributions of the \Upp beams at 252 and 165~\MeVu, respectively, measured by the \Xe gas IC are shown in Figures~\ref{fig-U_dE}(e) and (f). 
Only the \dE distributions of 90+ were measured as the difference in \dE distributions between 90+ and 91+ is expected to be smaller at lower energies. 
The energy resolutions at 252 and 165~\MeVu are both 1.4\%, enabling \Z separation in the uranium region. 
Although the energy resolution at 165~\MeVu was anticipated to improve more than at 252~\MeVu due to the larger \dE and \Ncc, no improvement was observed. 
This is attributed to columnar recombination occurring at high \dEdx conditions, as discussed in Section~\ref{sec6}.
}

\begin{table*}[htbp]
\centering
\caption{\label{tab-U_dE}
	Results of the \dE distributions of the \Up beams and calculated average number of charge-state changes (\Ncc).
	The required energy resolution is 2.1\% for \Z separation in the uranium region (see text).
}
\smallskip
\begin{tabular}{llllll}
   \hline
   IC gas   & \Up energy  & Mean \dE  &Mean \dE difference& Energy resolution & \Ncc\\
            & [MeV/u] &   90+ / 91+ [arb.unit]&[\%]& 90+ / 91+ [\%] &\\
   \hline
   P-10   &  344 &186.42 / 187.57& 0.62 &  3.0 / 2.9& 1.7\\
             & 252  &221.54 / 222.54& 0.45 & 2.2 / 2.7& 2.8 \\
             & 165  &289.05 / 289.40& 0.12 & 1.9 / 2.1& 7.8 \\
\Xe       & 344  &406.70 / 406.83& 0.03 & 1.6 / 1.6 & 23 \\
             & 252  &473.59 / -& - & 1.4 / -& 53 \\
             & 165  &601.57 / -& - & 1.4 / -& 175 \\
   \hline
\end{tabular}
\end{table*}

\newclearpage
\subsection{Cocktail RI beams of $Z=$40--90}

\begin{figure*}[p]
    \centering
	\Fig{\includegraphics[width=0.7\textwidth]{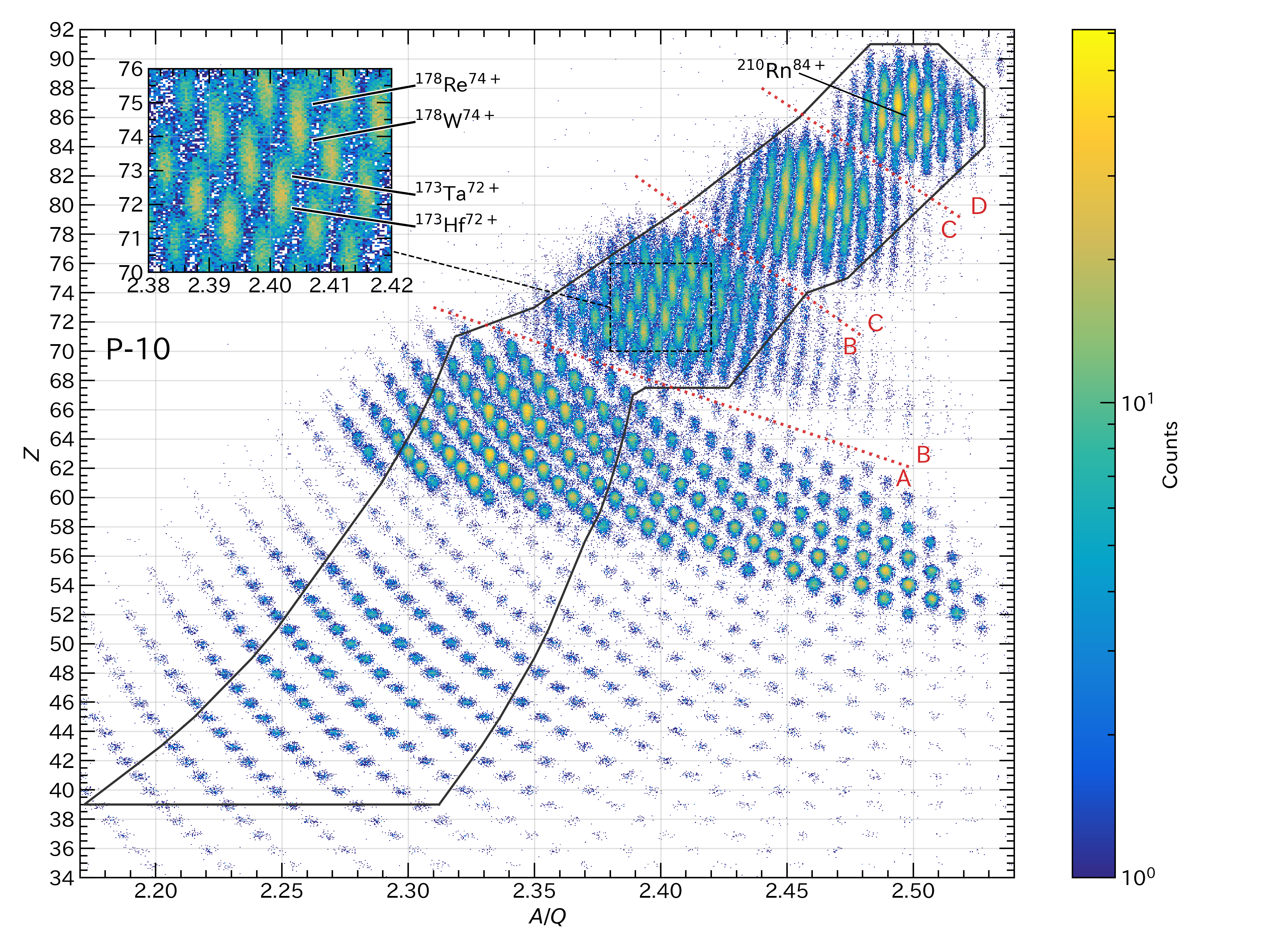}
    \\
    \includegraphics[width=0.7\textwidth]{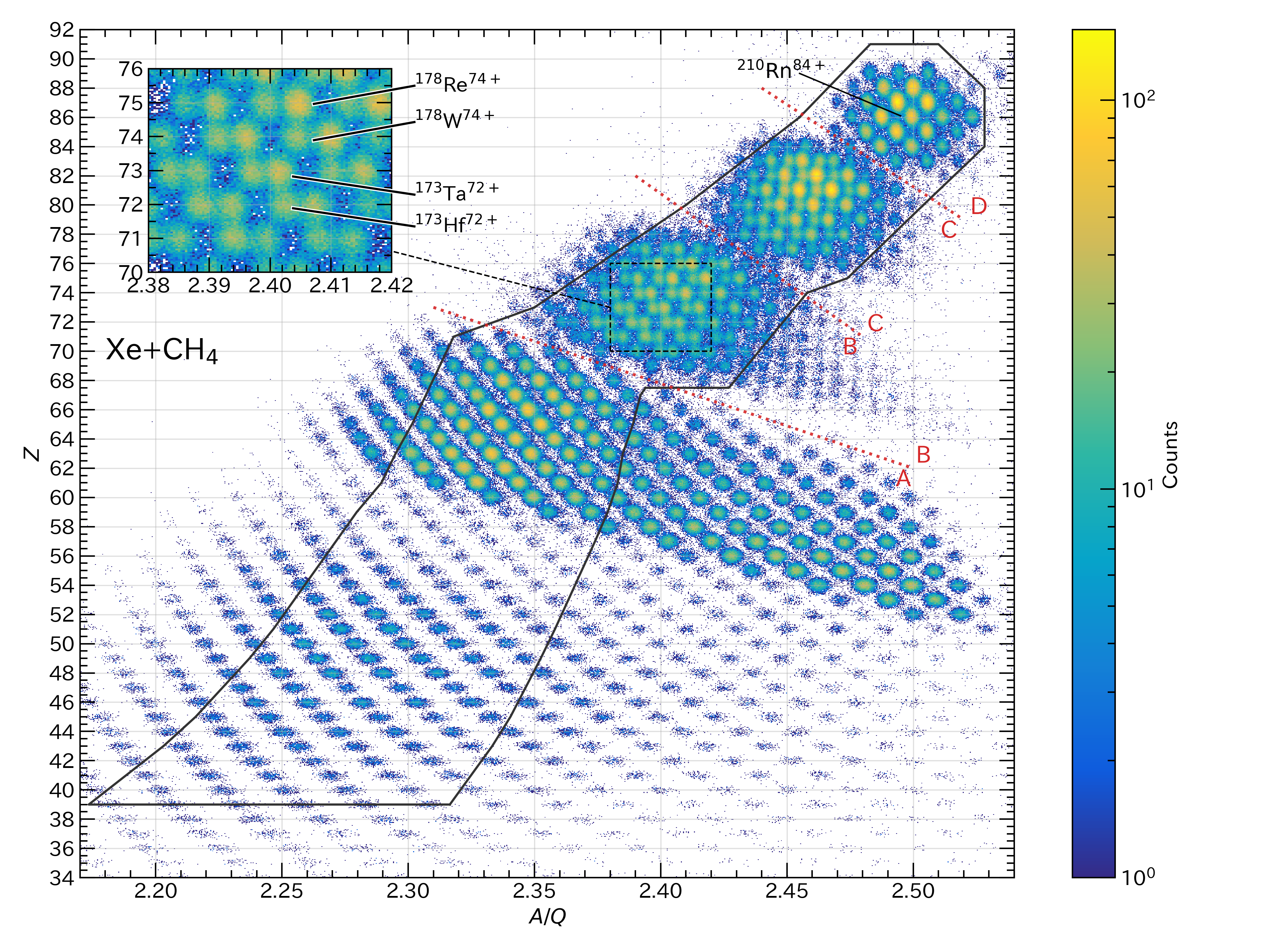}}
	\caption{
	PID plots of \Z vs. \AoQ for the cocktail RI beams with \Zeq{40}--90 injected into the P-10 (top) and \Xe (bottom) gas ICs.
	The beam energy is 200--250~\MeVu and the particle on the central trajectory is \Nuc{210}{86}{Rn}{84+}.
	The red dotted lines show the boundaries of groups A, B, C, and D according to the charge states of the ions.
	The charge states in groups A, B, C, and D are fully stripped, a mixture of fully stripped and H-like, a mixture of H-like and He-like, and He-like ions, respectively.
	The charge state of the ions in each cluster was identified using the condition that $A$, \Q, and \Z were integers (see text).
	The black solid lines indicate two-dimensional gates for evaluation of the beam energy and \Z resolution.
	}
	\label{fig-PID}
\end{figure*}

\English{
The \Z vs. \AoQ PID plots for the cocktail RI beams of \Zeq{40}--90 obtained with the P-10 and \Xe gas ICs are depicted in Fig.~\ref{fig-PID}. 
Ions with different \Z, $A$, and \Q are distributed as clusters in the PID plots, where \Q represents the charge state between F3 and F7. 
The \AoQ width of each cluster, \AoQ resolution, obtained with the \Xe gas IC (bottom) is inferior to that with the P-10 gas IC (top) due to poor TOF resolution caused by radiation damage to the plastic scintillator used for measurements with the \Xe gas IC. 
The TOF resolutions of 140~ps and 170~ps were obtained for the P-10 and \Xe gas ICs, respectively. 
This difference is negligible for \Z resolution considering error propagation from $\beta$ to \dE.
Subsequently, ions enclosed by the black solid lines are subject to analysis. 
}

\begin{figure}[htbp]
	\centering
	\Fig{\includegraphics[width=0.5\textwidth]{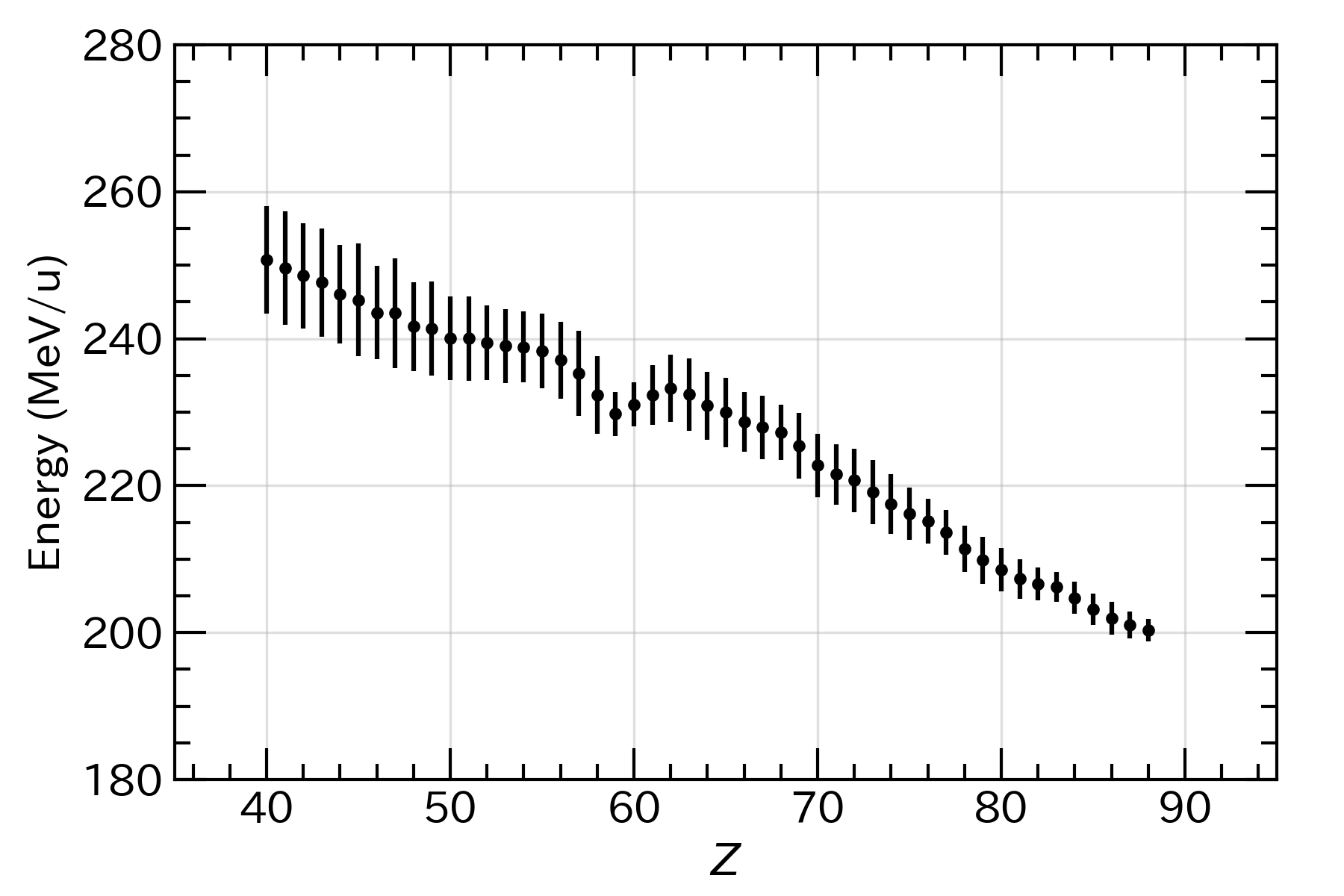}}
	\caption{
		Incident energy of the cocktail RI beams of \Zeq{40}--90 inside the two-dimensional gate in Fig.~\ref{fig-PID} as a function of \Z.
		The incident energy at the entrance of the F7 chamber was calculated from $\beta_{57}$.
		The error bars indicate the standard deviation (SD).
		The mean energy is 250.7 (7.3 in SD) \MeVu at \Zeq{40} and 200.4 (1.5 in SD) \MeVu at \Zeq{88}.
	}
	\label{fig-Z3090_energy}
\end{figure}

\English{
Figure~\ref{fig-Z3090_energy} shows the incident energy at the entrance of the F7 chamber, calculated from $\beta_{57}$, as a function of \Z.
The mean energy decreases with increasing \Z due to a larger energy loss for higher \Z ions in the F1 energy degrader.
The mean energy is 250.7~\MeVu at \Zeq{40} and 200.4~\MeVu at \Zeq{88}.
}

\English{
The charge state of the ions in each cluster was identified using the condition that $A$, \Q, and \Z were integers.
The ions in Group A were identified as fully-stripped ions because multiplying \AoQ by \Z resulted in integers.
For Group B ions, clusters existed for which multiplying \AoQ by \Z resulted in integers and nonintegers.
The former clusters were identified as fully-stripped ions and the subsequent clusters were identified as H-like ions because multiplying \AoQ by ($Z-1$) resulted in integers.
Fully-stripped ions \Nuc{A}{}{Z}{Z+} are adjacent to H-like ions \Nuc{A}{}{(Z\!+\!1)}{Z+} along the \Z axis, as shown in the inset in Fig.~\ref{fig-PID} of the \Xe gas IC.
Because the pair of clusters, \Nuc{A}{}{Z}{Z+} and \Nuc{A}{}{(Z\!+\!1)}{Z+} in the P-10 gas IC cannot be separated, the clusters were considered equivalent to those in the \Xe gas IC.
Similarly, groups C and D were identified.
Group C consists of H-like and He-like ions, whereas Group D comprises He-like ions.
}

\begin{figure}[htbp]
	\centering
	\Fig{\includegraphics[width=0.5\textwidth]{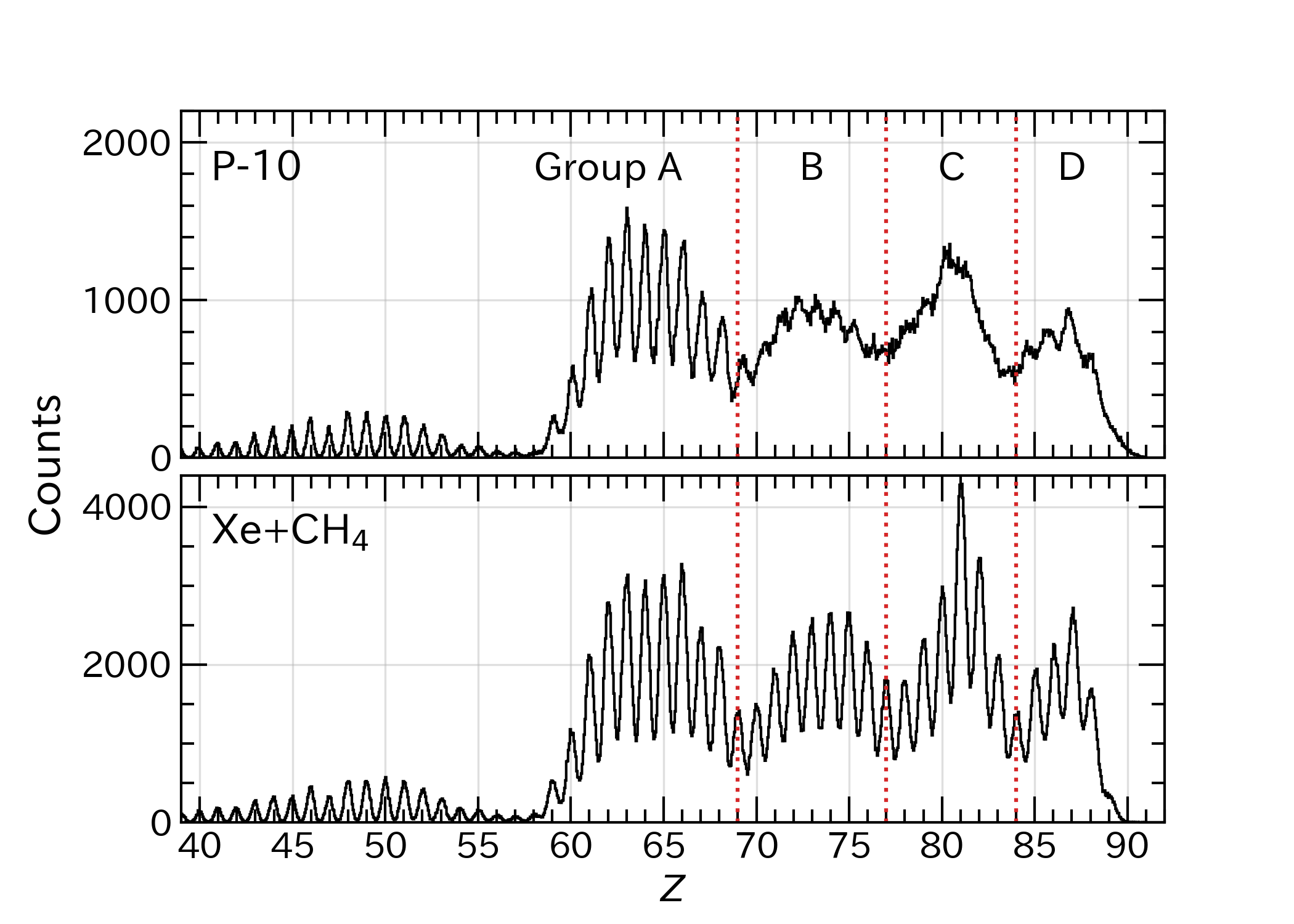}}
	\caption{
		\Z spectra of the cocktail RI beams inside the two-dimensional gates in Fig.~\ref{fig-PID} obtained with the P-10 and \Xe gas ICs.
		}
	\label{fig-Z_All}
\end{figure}

\English{
Figure~\ref{fig-Z_All} displays the \Z spectra. 
In the region of Group A, the \Z peaks of the P-10 gas IC closely resemble those of the \Xe gas IC. 
However, in the regions of groups B, C, and D, notable differences emerge: the \Z peaks of the P-10 gas IC are lower and less distinct, whereas those of the \Xe gas IC are separated up to \Zeq{88}.
}

\begin{figure}[htbp]
    \centering
	\Fig{
	\subfloat{\includegraphics[width=0.5\textwidth]{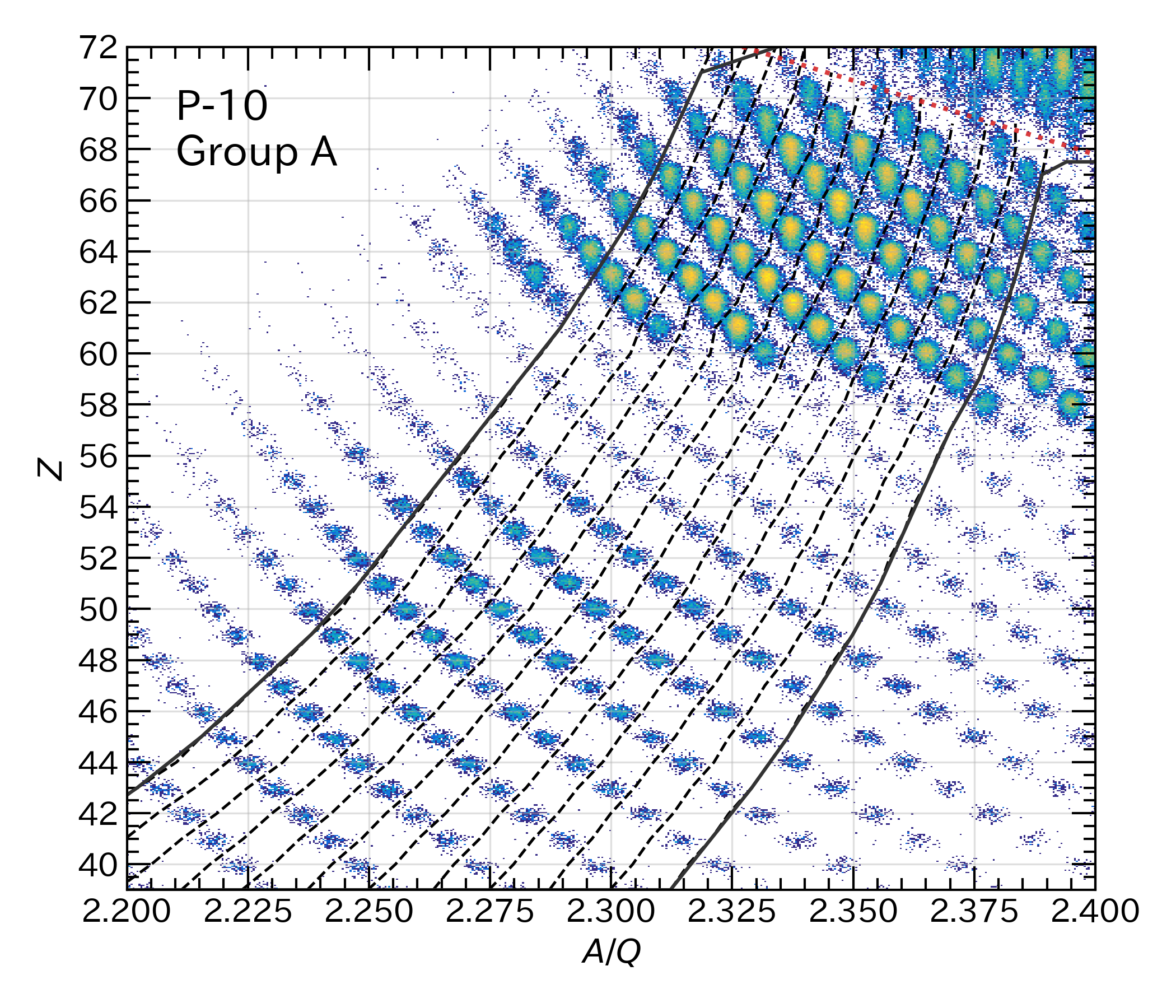}}
    \subfloat{\includegraphics[width=0.5\textwidth]{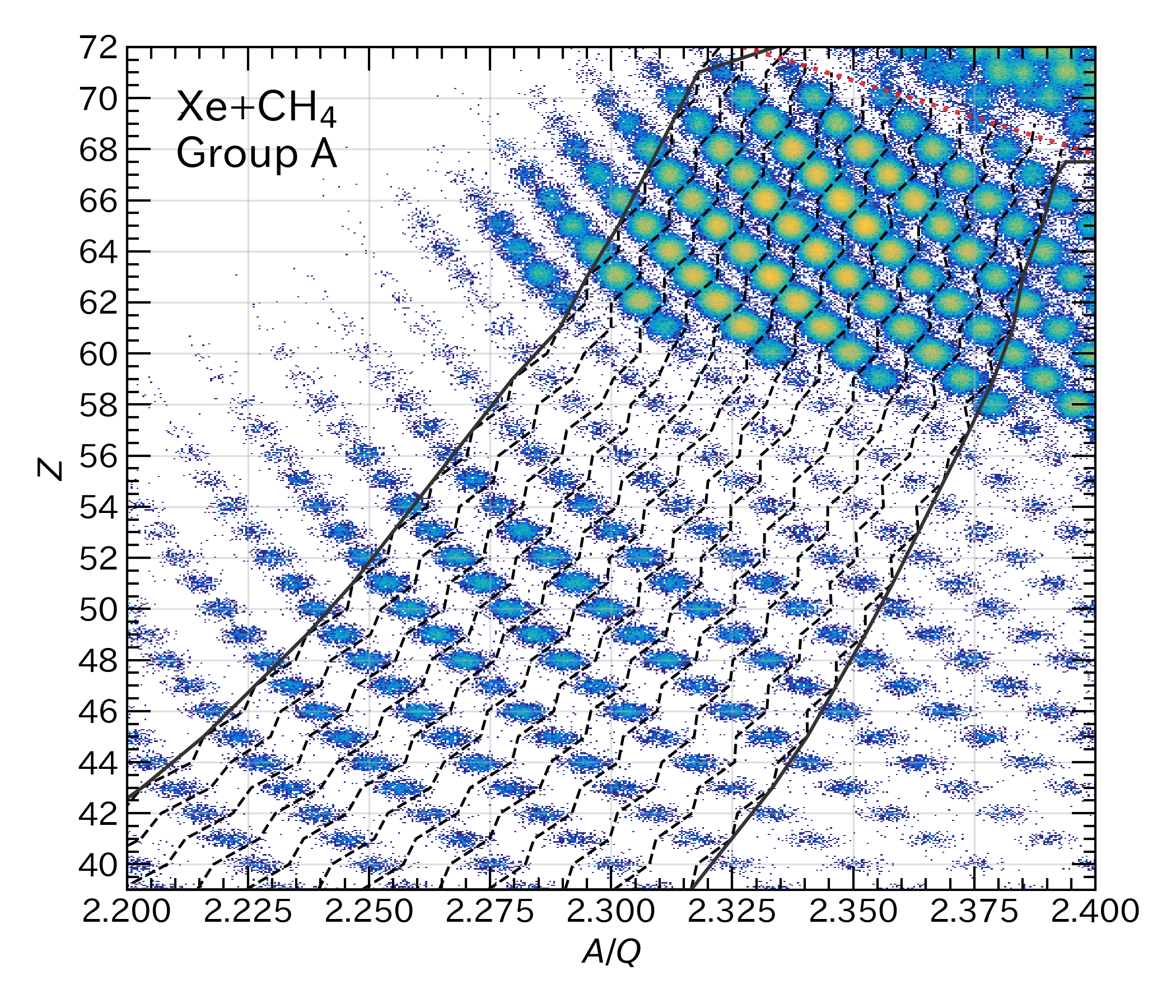}}
	}
	\caption{
	Example of ribbon-shaped gates indicated by dashed lines to divide the ions into even or odd \Z for Group A of the cocktail RI beams.
	Gates for Group D are similar to those for Group A.
	}
	\label{fig-PID_ribbon}
\end{figure}

\begin{figure}[htbp]
	\centering
	\Fig{
	\subfloat{\includegraphics[width=0.5\textwidth]{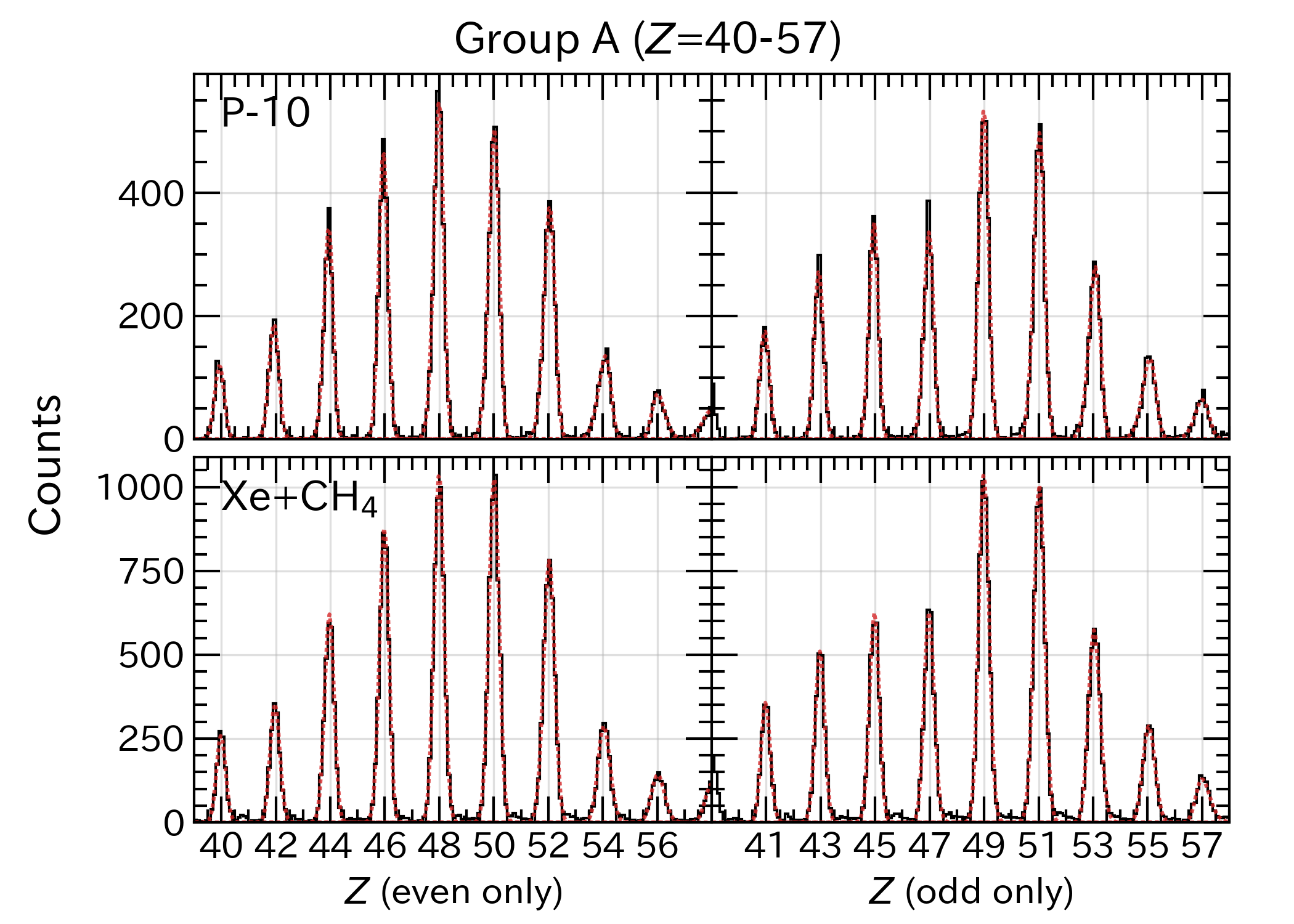}}
	\subfloat{\includegraphics[width=0.5\textwidth]{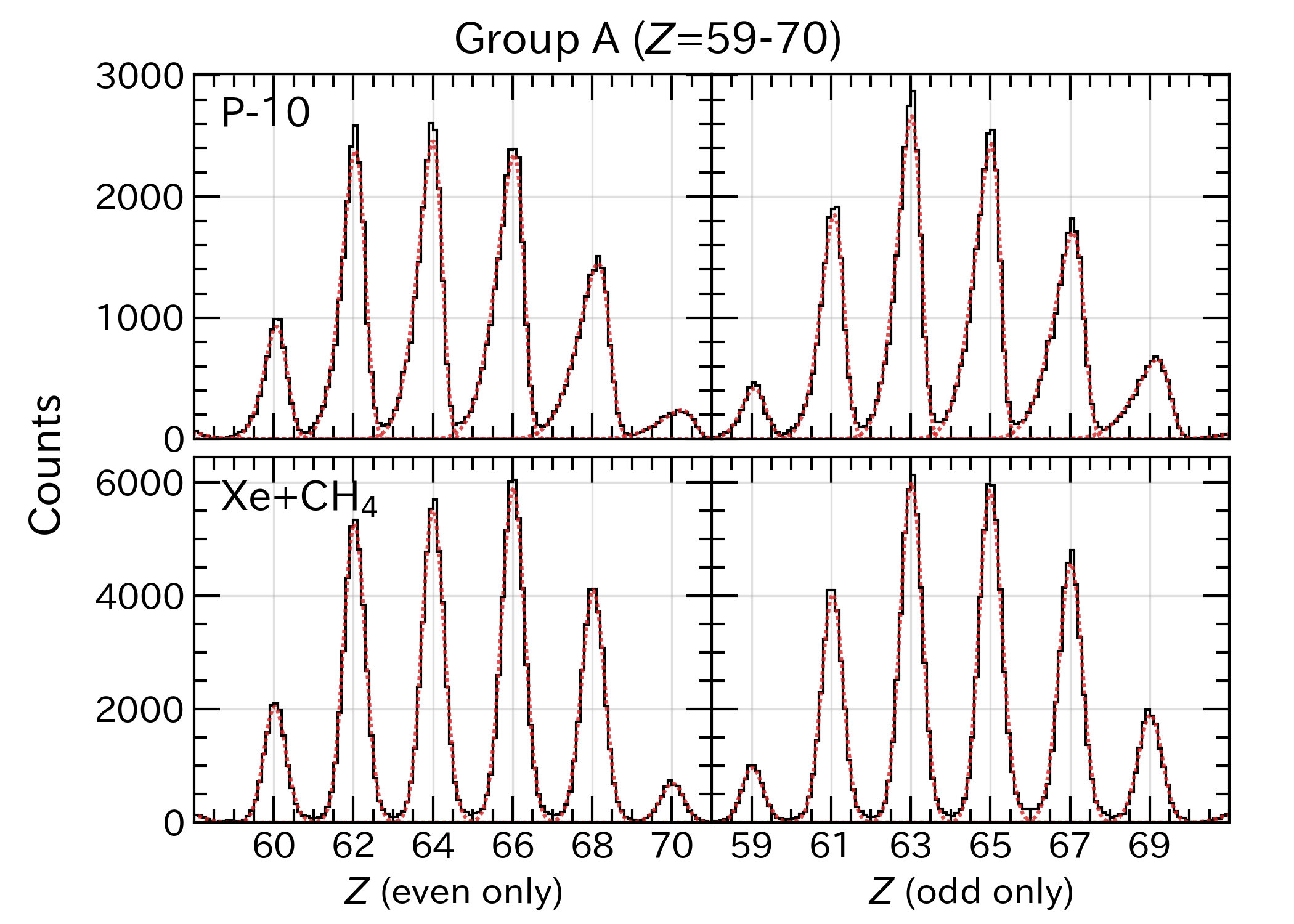}}
	}
	\caption{
		\Z spectra of even- (left) or odd-\Z (right) ions for Group A of the cocktail RI beams.
		The black solid lines show experimental data and the red dotted lines indicate the fitted skew-normal distributions.
		}
	\label{fig-Z_Aall}
\end{figure}

\begin{figure}[htbp]
	\centering
	\Fig{\includegraphics[width=0.5\textwidth]{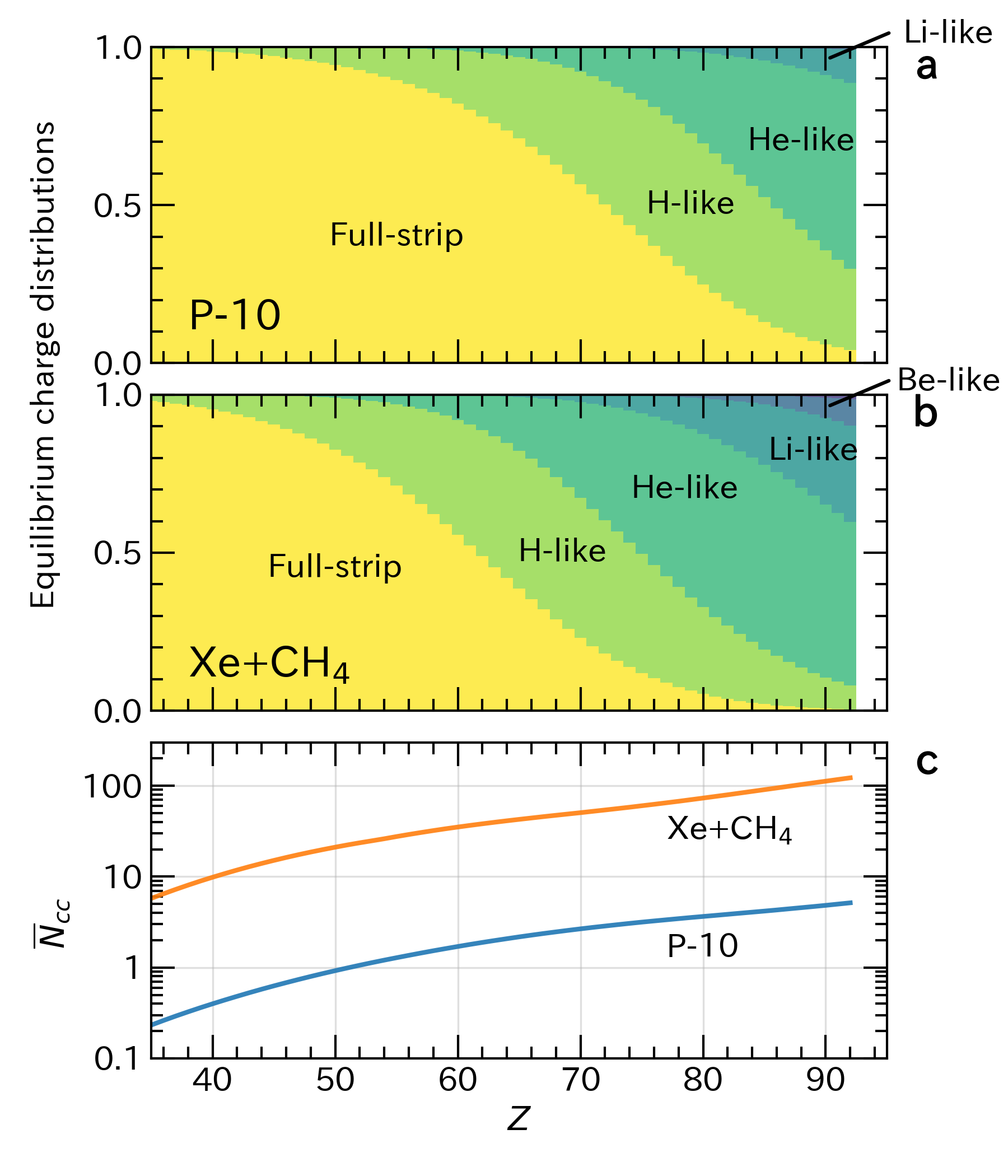}}
	\caption{
		Calculated equilibrium charge distributions of the cocktail RI beams and the calculated average number of charge-state changes of the beam at the equilibrium charge state in the IC gas (\Ncc).
		The charge-state changing cross section is obtained from the GLOBAL code~\cite{GLOBAL}.
	}
	\label{fig-Q_Ncs_e1}
\end{figure}

\English{
For the assessment of \Z resolution in Group A, ribbon-shaped gates separating even- and odd-\Z ions were applied (Fig.~\ref{fig-PID_ribbon}) to reveal tails hidden by neighboring peaks in the \Z spectra. 
The black solid lines in Fig.~\ref{fig-Z_Aall} show the \Z spectra of even- or odd-\Z ions with exposed tails. 
The \Z peaks of the \Xe gas IC are symmetrical, whereas those of the P-10 gas IC at \Zeq{59}--70 have low-\Z tails. 
In both gases, the most probable charge state is fully stripped, with the fraction of H-like ions increasing with \Z as shown in Fig.~\ref{fig-Q_Ncs_e1}. 
For the P-10 gas, the \Ncc of 1.6--2.7 is small to reach charge-state equilibrium, resulting in the large charge-exchange straggling.
The ions changing to H-like ions create low-energy tails in the \dEchex distribution, which also influence the \Z distribution. 
Conversely, in the case of the \Xe gas, the charge-exchange straggling is negligible for the collisional straggling because the \Ncc of 34--51 is sufficiently large. 
Consequently, symmetrical \Z distributions are observed. 
The red dotted lines in Fig.~\ref{fig-Z_Aall} represent fitted multiple skew normal functions representing tail components. 
The \Z resolutions of the P-10 and \Xe gas ICs are 0.45 and 0.44 at \Zeq{50}, 0.69 and 0.64 at \Zeq{62}, and 1.02 and 0.71 at \Zeq{69}, respectively. 
The rapid deterioration of the \Z resolution of the P-10 gas IC is attributed to the increased charge-exchange straggling.
}

\begin{figure}[htbp]
	\centering
	\Fig{\includegraphics[width=0.5\textwidth]{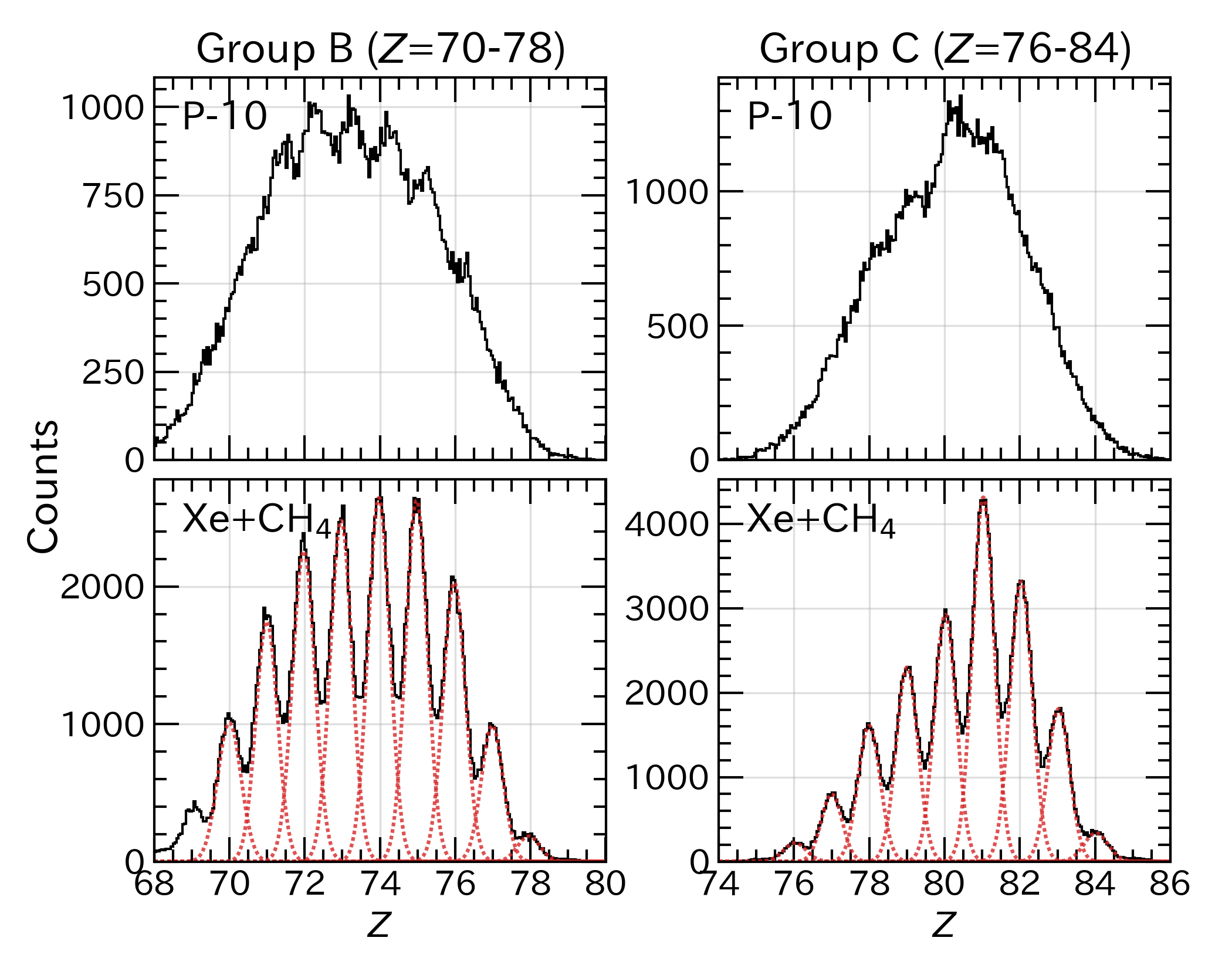}}
	\caption{
		\Z spectra of all \Z ions in groups B (left) and C (right) of the cocktail RI beams.
		The black solid lines show experimental data and the red dotted lines indicate the fitted normal distributions.
		The \Z spectra measured by the P-10 gas IC could not be fitted due to the unclear \Z peaks.
	}
	\label{fig-Z_BC}
\end{figure}

\English{
For groups B and C, separating even- and odd-\Z ions was not possible owing to the dense PID spectrum. 
Figure~\ref{fig-Z_BC} shows the \Z spectra of all \Z ions. 
The evaluation of \Z resolution for the P-10 gas IC was not completed as the fully-stripped \Nuc{A}{}{Z}{Z+} and H-like \Nuc{A}{}{(Z\!+\!1)}{Z+} ions, having the same \AoQ value and different \Z values, could not be separated. 
This difficulty is attributed not only to poor \Z resolution but also to the mean \dEchex dependence on the incident charge states, similar to the \Up-beam results of the P-10 gas IC. 
In the case of the \Xe gas IC, symmetrical \Z peaks were observed. 
The \Z distribution was fitted with multiple normal functions. 
The \Z resolutions of the \Xe gas IC were 0.69 at \Zeq{74} in Group B and 0.68 at \Zeq{81} in Group C. 
The \Z resolutions of the \Xe gas IC were almost constant within \Zeq{70}--84.
}

\begin{figure}[htbp]
	\centering
	\Fig{\includegraphics[width=0.5\textwidth]{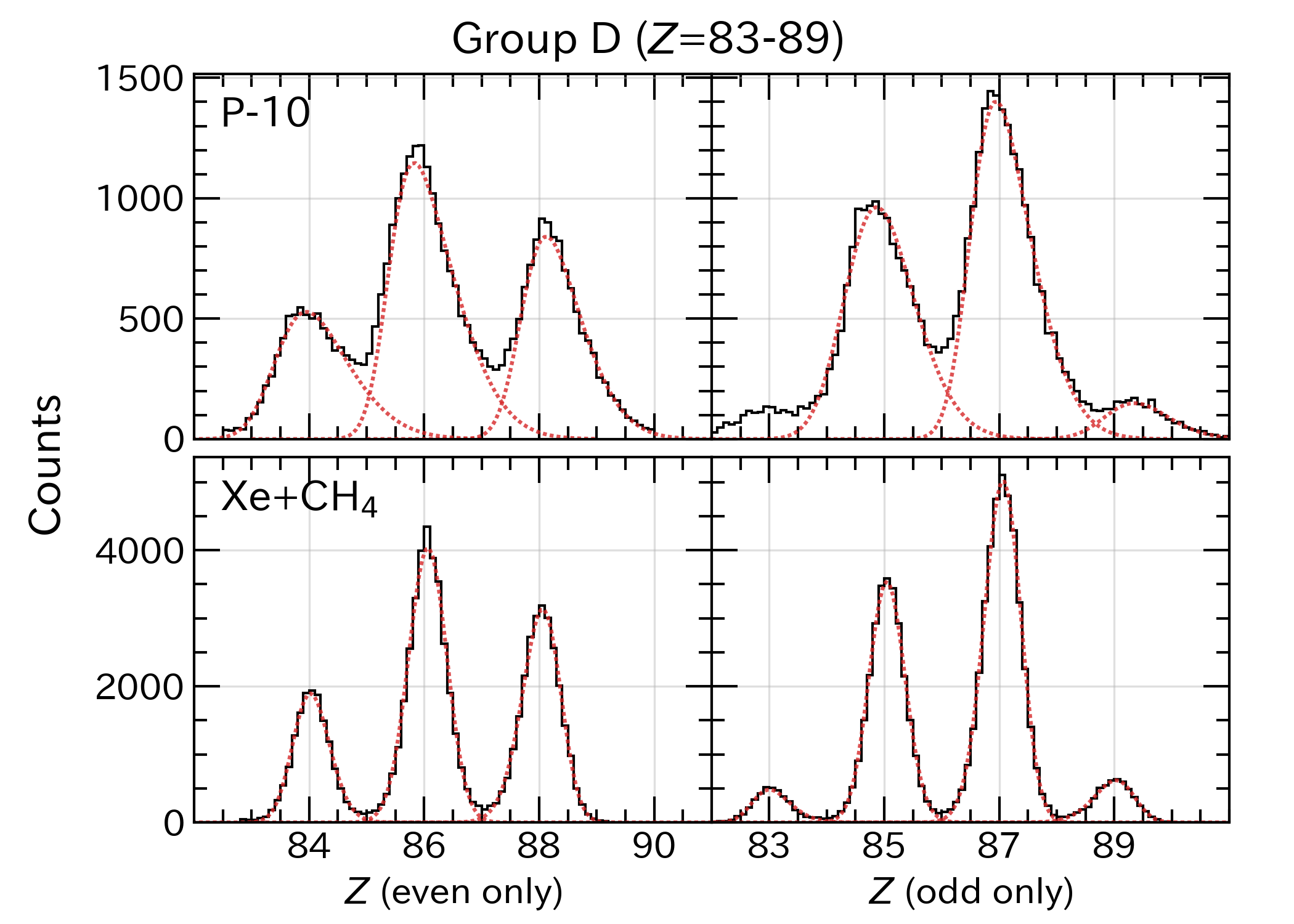}}
	\caption{
		\Z spectra of even- (left) and odd- (right) \Z ions for Group D of the cocktail RI beams.
		The black solid lines show experimental data and the red dotted lines indicate the fitted skew-normal distributions.
		}
	\label{fig-Z_Dall}
\end{figure}

\English{
For Group D, consisting of He-like ions, \Z resolution was evaluated as for Group A. 
The \Z spectra of even- and odd-\Z ions were separately analyzed (Figure~\ref{fig-Z_Dall}). 
The high-\Z tails of the P-10 gas IC were evaluated using multiple skew normal functions. 
The average \Z resolutions of the P-10 and \Xe gas ICs were 1.28 and 0.74 at \Zeq{84}--88, respectively. 
The worse \Z resolution of the P-10 gas IC is attributed to the increased charge-exchange straggling.
}

\begin{figure}[htbp]
	\centering
	\Fig{\includegraphics[width=0.5\textwidth]{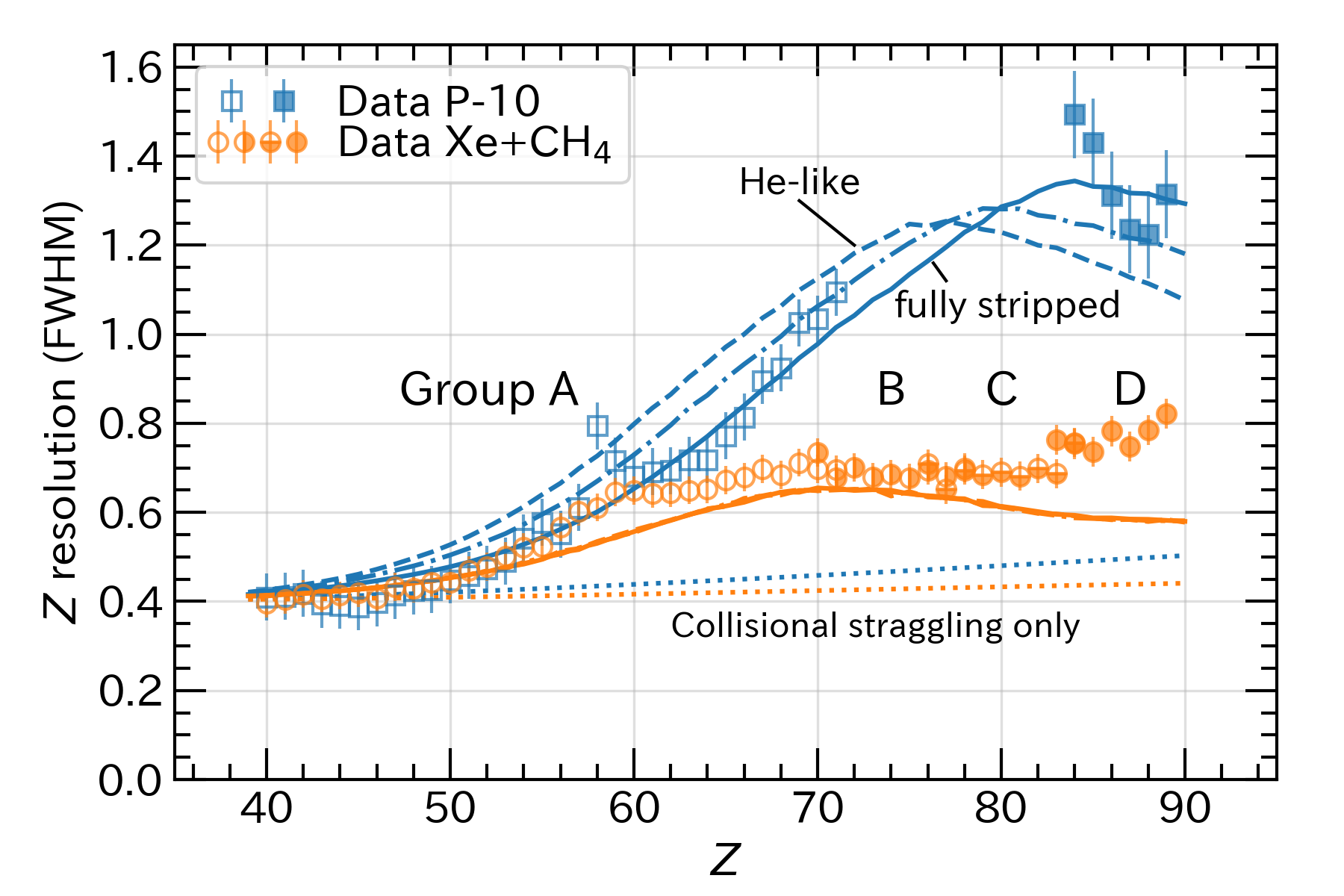}}
	\caption{
		\Z resolution of the cocktail RI beams with the P-10 and \Xe gas ICs.
		The blue squares and orange circles represent the measured \Z resolution (\Rexp) of the P-10 and \Xe gas ICs, respectively.
		The filled styles of the symbols indicate groups A--D.
		The blue and orange lines depict the simulated \Z resolution ($R_{sim}$)  of the P-10 and \Xe gas ICs, respectively.
		The dotted lines illustrate the \Z resolution (\Rcol) assuming only collisional straggling, with the truncated energy of the \drays adopted to reproduce \Rexp of \Zeq{40}--50.
		The other lines depict the \Z resolution including both collisional and charge-exchange straggling (\Rchex).
		The solid, dashed-dotted, and dashed lines denote incident ions with fully-stripped, H-like, and He-like, respectively.
		The three lines for the \Xe gas IC overlap each other due to nearly identical \Z resolutions.
	}
	\label{fig-Zres_Data_Sim_e}
\end{figure}

\English{
The \Z resolutions of all the groups are summarized in Fig.~\ref{fig-Zres_Data_Sim_e}. 
The blue squares and orange circles denote the data obtained by the P-10 and \Xe gas ICs, respectively. 
Both \Z resolutions at \Zl{60} were similar, while the \Z resolutions of the \Xe gas IC at \Zg{60} were better than those of the P-10 gas IC. 
The \Z resolution of the P-10 gas IC exceeded one at \Zg{70}, making it difficult to separate the $\delta Z=1$ difference. 
In contrast, the \Z resolution of the \Xe gas IC was less than 0.8 and achieved a 3$\sigma$ \Z separation or better over a wide range of \Zeq{40}--90.
}

\newclearpage

\section{Discussion\label{sec6}}
\subsection{Comparison with \dE straggling model for \Z resolution}

\English{
The measured \Z resolutions (\Rexp) by the P-10 and \Xe gas ICs were compared using the \dE straggling model, which includes collisional straggling and charge-exchange straggling, as explained in Section~\ref{sec2}.
The simulated \Z resolution (\Rsim) was derived from the simulated energy loss ($\Delta E_{sim}$) and the simulated \dE straggling ($\Omega_{sim}$) as follows: 
\begin{linenomath*}\begin{equation}\label{eq:zchex}
	\Delta Z_{sim}(Z) = \frac{2.355 \Omega_{sim}(Z)}{\Delta E_{sim}(Z)-\Delta E_{sim}(Z-1)},
\end{equation}\end{linenomath*}
where 2.355 is the ratio of the FWHM to standard deviation.
Note that \Rsim in this definition is equivalent to \Rexp.
}

\English{
Both \Rexp values by the P-10 and \Xe gas ICs were similar at \Zl{50}, although the \Z resolution of the \Xe gas IC is expected to be superior to that of the P-10 gas IC because of the double energy loss.  
Because the majority of the ions are fully stripped in the \Zl{50} region, charge-exchange straggling has a limited effect on \dE straggling.
To interpret the similar \Z resolutions in terms of the collisional straggling only, the truncated Bethe-Bohr model reported by Pf\"{u}tzner et al.\cite{Pfutzner1994NIMB} was used.
In the model, high-energy \drays escaping from the active detection volume are considered.
The decreased energy-loss rate ($dE_{colT}/dx$) and collisional straggling (\OBB) in the truncated Bethe-Bohr model are given by modifying Eqs.~\ref{eq:bethe} and \ref{eq:omega} as follows
\begin{linenomath*}\begin{equation}\label{eq:bethe2}
	\frac{d E_{colT}}{d x} =  Z^2\frac{4 \pi  e^4}{m\beta^2c^2}n_e\left(\frac{1}{2} \ln\frac{E_m E_d}{I^2} - \beta^2  +\Delta_{cor}\right)
\end{equation}\end{linenomath*}
and
\begin{linenomath*}\begin{equation}\label{eq:omega2}
	\frac{{d\Omega_{colT}^2}}{d x}  =  Z^2\frac{2 \pi  e^4}{m\beta^2c^2} n_e E_d X_{LS},
\end{equation}\end{linenomath*}
respectively, with truncated \dray energy, \Ed ($\le$\Em) defined in Ref.\cite{Pfutzner1994NIMB}, which corresponds to the maximum energy of \drays absorbed in the active detection volume.
The simulated \Z resolution (\Rcol) was derived from \dEBB and \OBB using Eq.~\ref{eq:zchex}.
}

\English{
The truncated \dray energy of \Ed was determined such that \Rcol reproduces the \Rexp at \Zeq{40}--50 for the cocktail RI beams in this study. 
For the \Xe gas IC, a value of \Ed equal to 620~keV was obtained. 
Given that the maximum \dray energy (\Em) is 620~keV for the 250-\MeVu cocktail RI beam at \Zeq{40}, it is assumed that the entire range of \drays was detected in the active detection volume. 
In the case of the P-10 gas IC, \Ed was determined to be 300~keV, approximately half of \Em. 
This twofold difference in \Ed between the \Xe and P-10 gas ICs can be attributed to the twofold difference in the energy-loss rate. 
In a previous study at RIBF~\cite{Sato2012APR}, \Ed of 310~keV was reported for cocktail RI beams of \Zeq{40}--50 at approximately 330~\MeVu, measured with a similar P-10 gas IC. 
However, the electrode size in the previous IC was 26$\times$17~cm$^2$, considerably larger than the present IC with a size of \diameter 6~cm.
Reference~\cite{Pfutzner1994NIMB} suggested that \Ed is independent of the beam energy; however, it is reasonable to consider that \Ed depends on the active detection volume. 
The similar \Ed values obtained by ICs with different electrode sizes could be due to the simplistic approach of this model, as pointed out by Pf\"{u}tzner et al.~\cite{Pfutzner1994NIMB}, or the complex active detector volume, including support rings and electrode films. 
To investigate the dependence on the electrode size, additional experiments are required in which the same RI beams are injected into two ICs with different-size electrodes. 
In Fig.~\ref{fig-Zres_Data_Sim_e}, the \Rcol values calculated up to \Zeq{90} using the obtained \Ed values are shown by the dotted lines. 
The \Rcol values were nearly independent of \Z. 
The \Rexp values at \Zg{50} are considerably larger than \Rcol in both ICs, indicating the effect of the charge-exchange straggling.
}

\English{
In the Monte Carlo simulation of the charge-exchange straggling, the IC and upstream PPAC are taken into consideration. 
The charge-state changing cross section is obtained from the GLOBAL code~\cite{GLOBAL}, considering charge states from fully stripped to C-like. 
The \dEchex distribution is derived from Eq.~\ref{eq:chex2}, assuming a constant $\beta$ in the gas. 
Simulated \Z resolution, \Rchex with collisional and charge-exchange straggling was derived using Eq.~\ref{eq:zchex} from \dEchex and the total \dE straggling ($\Omega_{total}$), which is the width of the \dEchex distribution convoluted with \OBB.
}

\English{
The \Rchex values with fully-stripped, H-like, and He-like incident ions are shown by the solid, dashed-dotted, and dashed lines in Fig.~\ref{fig-Zres_Data_Sim_e}, respectively.
There are two features of the P-10 gas IC. 
First, \Rchex depends on the incident charge states.
The cause of this dependence is the same as that of the difference in the \dE distributions of \Nuc{}{}{U}{90+,91+} beams, as shown in Fig.~\ref{fig-U_dE}.
The charge-state distributions in the P-10 gas IC depend on the incident charge state owing to the small \Ncc of $<$5 as shown in Fig.~\ref{fig-Q_Ncs_e1}(c).
Second, the incident charge state, which gives the best \Rchex depends on \Z.
The fully-stripped ions give the best \Rchex for $Z\leq78$, whereas the He-like ions give the best \Rchex for $Z\geq79$.
These preferable charge states approximately correspond to the most probable charge state, which is fully stripped for $Z\leq74$ and He-like for $Z\geq85$ as shown in Fig.~\ref{fig-Q_Ncs_e1}(a).
For the \Xe gas IC, the \Rchex values are independent of the incident charge states because of the large \Ncc of $>$20 at \Zg{50}, which is sufficiently large for the charge distribution to reach equilibrium near the upstream of the active gas.
}

\English{
The \Rchex lines in Fig.~\ref{fig-Zres_Data_Sim_e} generally reproduce the \Rexp values for both ICs: they degrade with increasing \Z, and their disparity between the P-10 and \Xe gas ICs amplifies at \Zg{50}. 
This implies that \Rexp is reasonably explained by the \dE straggling model. 
In more granular detail, the \Rexp values tend to be worse than the \Rchex in the high-\Z region for both ICs. 
For instance, the \Rexp values for the P-10 gas IC at \Zg{84} appear to be worse than the \Rchex line of the incident He-like ions, and those for the \Xe gas IC at \Zg{80} are inferior to the \Rchex lines. 
Similar phenomena are observed in the energy resolutions of the \Up beams for the \Xe gas IC. 
The energy resolution did not improve even as \Ncc increased with a decrease in the beam energy from 252 to 165~\MeVu, as described in Section~\ref{sec-ubeam}. 
One plausible explanation for these disparities is columnar recombination. 
Columnar recombination takes place near the trajectory of high \dEdx beams, where the density of low-energy ionization electrons and positive gas ions is extremely high. 
The recombination of the low-energy electrons and gas ions reduces the \dE value, leading to a decrease in the mean \dE difference between \Z and $Z-1$ ions. 
The measured \dE at \Zg{70} with the \Xe gas IC is smaller than the \dEchex values in Eq.~\ref{eq:chex2}, although the other measured \dE values with the P-10 and \Xe gas IC are nearly reproduced by the calculated ones. 
Given that the \dEdx with the \Xe gas IC is the largest in our measurements, this discrepancy could be attributed to columnar recombination. 
On the contrary, the collisional straggling is likely unaffected by columnar recombination, as high-energy \drays are spatially dispersed and less prone to recombine. 
Consequently, the \Z resolution and energy resolution are expected to qualitatively worsen under high \dEdx conditions.
}

\subsection{Perspective}

\English{
The IC gases alter the charge distribution of heavy ion beams after passing through the ICs; however, this change does not affect inverse-kinematic nuclear--reaction experiments at RIBF.
The beam position at F8, where secondary targets are located, remains independent of the charge state after F7 
because these two focal planes are connected via a straight beamline without dipole magnets~\cite{Kubo2012PTEP}.
}

\English{
Notably, the \Xe gas induces approximately two times the energy loss compared to the P-10 gas. 
For instance, the energy loss of \Nuc{210}{}{Rn}{} at 250~\MeVu is 6.4~\MeVu when using the active gas length of the P-10 gas, whereas it is 13.7~\MeVu with that of the \Xe gas. 
Despite this difference, it remains acceptable for most experiments conducted at RIBF. 
To further mitigate the energy loss induced by the \Xe gas IC, the gas pressure can be reduced. 
Even if the gas pressure of the \Xe is halved to achieve energy-loss equivalence with the P-10 gas IC, a sufficient \Ncc is maintained. 
This is due to the charge-state changing cross section of the \Xe gas being one order of magnitude higher than that for the P-10 gas. 
Although the reduction in \dE may lead to a degradation in \Z resolution, a decreased truncated \dray energy (\Ed) resulting from a lower energy-loss rate may compensate for the compromised \Z resolution.
}

\section{Summary}

\English{
The \Xe gas ionization chamber (IC) was developed to identify \Z of heavy ion beams with \Zg{70} at RIBF. 
To assess the energy and \Z resolutions of both the commonly-used P-10 gas and the newly-proposed \Xe gas ICs, \Nuc{238}{92}{U}{90+,91+} beams at 165, 252, and 344~\MeVu, and cocktail radioactive isotope (RI) beams with \Z ranging from 40 to 90 at 200--250~\MeVu were injected into these ICs. 
Results from the \Up beam injected into the P-10 gas IC indicated difficulty in \Z separation in the uranium region. 
This difficulty arises from a substantial difference in mean energy losses between \Nuc{}{}{U}{90+} and \Nuc{}{}{U}{91+} beams, coupled with an energy resolution of 1.9--3.0\% in FWHM, which is insufficient for the required 2.1\% for \Z separation in the uranium region. 
Conversely, results with the U beams in the \Xe gas IC demonstrated sufficient performance for \Z separation. 
The negligible mean energy-loss difference and an energy resolution of 1.4--1.6\% were better than the required resolution of 2.1\%. 
Results with the cocktail RI beams revealed that the \Xe gas IC exhibited a superior \Z resolution compared to the P-10 gas IC at \Zg{65}. 
The FWHM of the \Z resolution of the P-10 gas IC at \Zg{70} was larger than 1, rendering \Z separation challenging. 
In contrast, the \Z resolution of the \Xe gas IC was 0.69 at \Zeq{67}--70 and 0.74 at \Zeq{84}--88, proving sufficient for \Z separation. 
With the \Xe gas IC, 3$\sigma$ \Z separation or better was achieved in a wide \Z region up to \Zeq{90}. 
The difference in \Z resolution at \Zg{65} between the P-10 and \Xe gas ICs was explained by simulated charge-exchange straggling, in addition to collisional straggling using the truncated Bethe-Bohr model. 
The \Xe gas IC is suitable for \Z identification of heavy ion beams with multi-charge states in the energy range 200--300~\MeVu, without a significant increase in the energy loss in the IC gas thickness.
Therefore, the \Xe gas IC considerably facilitates the production of heavy RI beams in RIBF and other facilities.
}

\section{Acknowledgements}

The authors acknowledge the staff of the RI Beam Factory.
The experiments were conducted at the RI Beam Factory operated by RIKEN Nishina Center and the Center for Nuclear Study, University of Tokyo.
This work was supported by the JSPS A3 Foresight Program, ``Nuclear Physics in the 21st Century''.
The authors would like to thank S. Kukita for his support of our theoretical understanding.

\end{document}